\documentclass[sigplan,nonacm,screen]{acmart}
\usepackage{subcaption}
\usepackage{booktabs}
\usepackage{bbm}
\usepackage{array}
\usepackage{makecell}
\usepackage{algorithm}
\usepackage{algpseudocode}
\usepackage{hyperref}
\usepackage{enumitem}
\usepackage{tcolorbox}
\tcbset{colback=blue!5!white,colframe=blue!50!black,boxrule=0.5pt}
\usepackage{tabularx}
\usepackage{comment}
\usepackage{cuted}
\usepackage{diagbox}
\usepackage{balance}
\usepackage{multirow}

\newcommand{\para}[1]{\par\medskip\noindent\textbf{#1.}}

\usepackage{xspace}
\xspaceaddexceptions{'}
\newcommand{\ourscheme}{RTeAAL Sim\xspace}

\newcommand{\name}{\ourscheme: Using Tensor Algebra to Represent and Accelerate RTL Simulation (Extended Version)}

\newcommand{\layerinput}{LI}
\newcommand{\operationinputmask}{OIM}
\newcommand{\layeroutput}{LO}

\AtBeginDocument{%
  }

\begin{document}

\title{\name}


\author{Yan Zhu}
\orcid{0000-0002-6932-7971}
\affiliation{%
  \institution{University of California, Berkeley}
  \city{Berkeley}
  \state{CA}
  \country{USA}}
\email{zhuyan14@berkeley.edu}

\author{Boru Chen}
\orcid{0009-0001-8024-1116}
\affiliation{%
  \institution{University of California, Berkeley}
  \city{Berkeley}
  \state{CA}
  \country{USA}}
\email{boruchen@berkeley.edu}

\author{Christopher W. Fletcher}
\orcid{0000-0002-9995-5995}
\affiliation{%
  \institution{University of California, Berkeley}
  \city{Berkeley}
  \state{CA}
  \country{USA}}
\email{cwfletcher@berkeley.edu}

\author{Nandeeka Nayak}
\orcid{0009-0006-8895-8206}
\affiliation{%
  \institution{University of California, Berkeley}
  \city{Berkeley}
  \state{CA}
  \country{USA}}
\email{nandeeka@berkeley.edu}









\begin{abstract}
RTL simulation on CPUs remains a persistent bottleneck in hardware design.
State-of-the-art simulators embed the circuit directly into the simulation binary, resulting in long compilation times and execution that is fundamentally CPU frontend-bound, with severe instruction-cache pressure.

This work proposes \ourscheme, which reformulates RTL simulation as a sparse tensor algebra problem.
By representing RTL circuits as tensors and simulation as a sparse tensor algebra kernel, \ourscheme decouples simulation behavior from binary size and makes
RTL simulation amenable to well-studied tensor algebra optimizations. 
We demonstrate that a prototype of our tensor-based simulator, even with a subset of these optimizations, already mitigates the compilation overhead and frontend pressure 
and achieves performance competitive with the highly optimized Verilator simulator across multiple CPUs and ISAs.
\end{abstract}






\maketitle

\section{Introduction}
\label{sec:intro}

Efficient development and verification of hardware designs requires fast and accurate Register-Transfer-Level (RTL) simulation.
While hardware-accelerated RTL simulation tools exist, they are expensive to acquire~\cite{palladium} 
and/or incur long compilation times for each new design~\cite{firesim}.
As a result, software-based RTL simulators that run on CPUs~\cite{essent, repcut, dedup, event-driven1, event-driven2, verilator1} remain widely used in both academia and industry.

Modern high-performance RTL simulators~\cite{essent,verilator1, repcut, dedup} 
for CPUs
translate hardware designs expressed in hardware description languages (HDLs) into nearly straight-line C++ programs, which are compiled into executable binaries using off-the-shelf C++ compilers. 
This approach introduces inefficiencies 
at both compilation and simulation time.

During compilation, increasing the design's size leads to more generated C++ code and a larger eventual binary.
Large C++ programs require substantial compilation time and memory, especially under aggressive optimization settings (e.g., \texttt{clang -O3}).

During simulation, the resulting binary exhibits low code reuse, placing heavy pressure on the instruction cache and creating frequent processor stalls at instruction fetch. 
This frontend bottleneck is fundamental, as these simulators rely on straight-line code to maximize the effectiveness of conventional compiler optimizations.

Conversely, tensor algebra expresses computation through a compact loop-based representation, offering an alternative to large, statically-generated instruction sequences. 
The resulting smaller binaries naturally reduce compilation cost while improving code reuse, which in turn alleviates instruction-cache pressure during simulation.
Moreover, the tensor algebra community has developed a rich body of well-established techniques, such as tensor compression and loop transformations.
These optimizations further reduce instruction footprint, improve memory locality and data reuse, and balance instruction- and data-cache pressure ~\cite{dragon-book, opt-compiler, cache-blocking, Iterative, taco, teaal, csf, format-abstraction, sze:2020:epo}. 

\para{This work}
Following the above observations, we propose \emph{\ourscheme}---a reformulation of RTL simulation as a tensor algebra problem. 
The core idea is to express the RTL dataflow graph as a sparse tensor and 
its execution as a \emph{cascade of extended Einsums}—a sequence of tensor algebraic equations involving the dataflow graph tensor and other tensors, augmented with user-defined operators~\cite{edge}. 

To optimize the \ourscheme formulation, we leverage TeAAL~\cite{teaal}, a framework for analyzing, scheduling, and optimizing sparse tensor algebra kernels.
TeAAL employs a methodology based on a separation-of-concerns that distinguishes the cascade (algorithm), mapping (dataflow), format (tensor 
layout), and binding (hardware allocation).
In this work, we primarily focus on optimizing sparse tensor storage formats (format level)~\cite{format-abstraction} and code transformations like loop unrolling~\cite{openequivariance} (binding level).
To highlight the generality and adaptability of \ourscheme, we map prior RTL simulation optimizations to 
TeAAL's abstractions.
In sum, we show that \ourscheme ameliorates traditional bottlenecks in RTL simulation,
preserves bespoke optimizations used in RTL simulators today and enables new opportunities for optimization.

To demonstrate that \ourscheme is practical, we implement and evaluate a simulator that incorporates its ideas.  
The simulator takes an RTL design described in FIRRTL and generates the corresponding tensors and a sparse tensor algebra kernel for performing RTL simulation.
Although this simulator is a proof-of-concept and implements only a subset of potential optimizations, we show that it alleviates bottlenecks from prior work
and is simulation performance-competitive with the popular simulator Verilator.
These results demonstrate the potential of our approach.

To summarize, we make the following contributions: 
\begin{itemize}
    \item We propose \ourscheme, a reformulation of RTL simulation as a tensor algebra problem that can 
    incorporate a wide range of tensor algebra optimizations.
    \item We use TeAAL to optimize the \ourscheme formulation.
    To demonstrate generality, we taxonomize bespoke RTL simulation optimizations in terms of TeAAL’s separation of concerns.
   \item We design and evaluate a concrete instance of \ourscheme that leverages a subset of 
   potential optimizations. 
   This proof-of-concept simulator achieves performance that is competitive with Verilator across
   four host machines.
   We open-source our codebase on GitHub\footnote{
   https://github.com/TAC-UCB/RTeAAL-Sim
   }.
\end{itemize}
\section{Background}
\label{sec:background}

\subsection{RTL Simulation}
\label{sec:background:rtl-sim}


RTL simulation models the behavior of digital circuits at the 
cycle level and captures how data flows between registers and combinational logic (CL).
During simulation, inputs specified in testbenches are applied to the design, and signal values are computed and propagated through the circuit.

\begin{figure}[h]
\centering
\includegraphics[width=\linewidth]{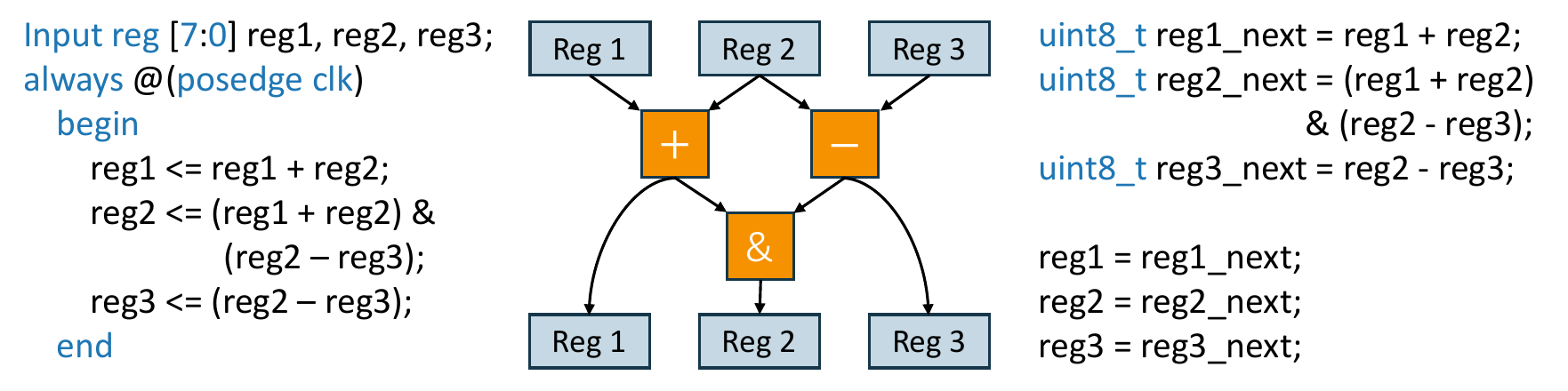} 
\caption{\textbf{Workflow for CPU- and compilation–based RTL simulation. The RTL of a synchronous circuit (left) is lowered onto a dataflow graph (middle) capturing the next-state logic. The simulator then compiles this graph into explicit C++ code (right) to compute and commit register next states each cycle.}
}
\label{fig:background:df-graph}
\end{figure}

As Figure~\ref{fig:background:df-graph} shows, software simulators typically compile the RTL design (left) into a \emph{dataflow graph} (middle), whose nodes represent primitive operations, and edges represent data flow.
This graph is then translated into C++ (right) for compilation and execution. 

RTL simulators can broadly be classified into two paradigms based on activity-awareness: \emph{event-driven}~\cite{rtl-sim-hw-sw-codesign, icarus-verilog, event-driven1, event-driven2, event-driven3} or \emph{full-cycle}~\cite{essent, essent-mag, repcut, dedup, parendi, gem} simulation.
Event-driven simulators dynamically schedule evaluation of nodes in the dataflow graph when inputs to those nodes are updated; 
i.e., they are \emph{activity-aware}.
In contrast, full-cycle simulators produce a static schedule at compile time, which is used to evaluate the circuit at runtime; i.e., they are \emph{activity-oblivious}.
Full-cycle simulators often perform better 
because they eliminate the overheads associated with tracking 
which nodes need to be re-evaluated~\cite{essent, essent-mag}.
For this reason, we focus on full-cycle simulation in this work.

\subsection{Tensors and Fibertrees}
\label{sec:background:tensor}

\begin{figure}[t]
\centering
\includegraphics[width=\linewidth]{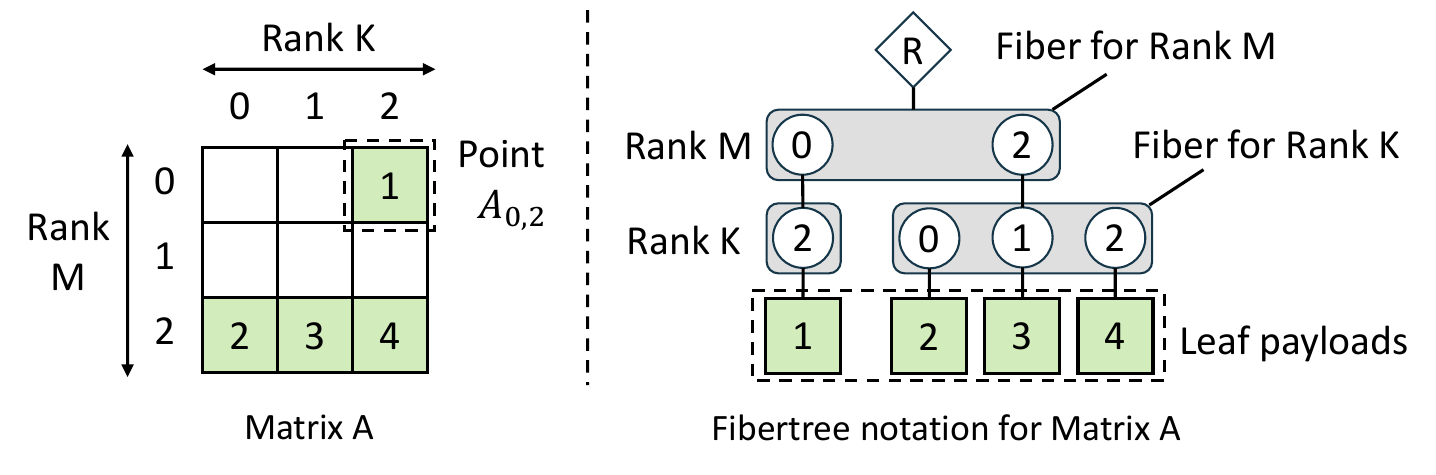} 
  \caption{\textbf{A matrix $A$ and its fibertree representation.
The matrix has two ranks, row rank $M$ and column rank $K$. $A_{0,2}$ is an element at point $(0, 2)$ with scalar value $1$. In the fibertree view, rank $M$ contains a single fiber of shape $3$ with occupancy $2$, while rank $K$ contains two fibers with occupancies $1$ and $3$, respectively, both with shape $3$. 
The leaf payloads correspond to the scalar values stored in the matrix.
  }}
  \label{fig:background:fibertree}
\end{figure}
In this paper, we adopt the tensor algebra notation defined in Sze et al.~\cite{sze:2020:epo}:

\begin{itemize}
\item \emph{Rank}: an axis/dimension of a tensor. For example, a matrix has two ranks (rows and columns).
\item \emph{Point}: a location in a tensor with a scalar value, identified by a tuple of coordinates. For example, $A_{m,k}$ denotes tensor $A$’s element at point $(m,k)$.
\end{itemize}

Mathematically, tensors have no notion of sparsity or compression format (e.g., CSR).
To avoid getting bogged down in the details of various formats, we use the following abstractions proposed in Sze et al.~\cite{sze:2020:epo} (see Figure~\ref{fig:background:fibertree} for illustration):

    
    
    
    

\begin{itemize}
\item \emph{Fibertree}: a tree representation of a tensor with levels corresponding to ranks. Each level contains \emph{fibers}.
\item \emph{Fiber}: a set of \emph{(coordinate, payload) pairs} sharing higher-level coordinates.
\item \emph{Payload}: a scalar value when it is leaf or a reference to the next-level fiber when it is an intermediate node.
\item 
\emph{Shape}: the number of all possible coordinates in a fiber, regardless of whether those coordinates correspond to empty payloads.
The shape of a rank equals that of its fibers.
We use the same symbol (e.g., $M$) for the name and shape of a rank.
\item \emph{Occupancy}: the number of coordinates with non-empty payloads in a fiber.
\end{itemize}


One advantage of fibertrees is that they naturally handle both dense and sparse tensors.
A dense tensor's fibertree explicitly contains all coordinates in the shape, while a sparse tensor's fibertree omits coordinates with empty payloads.




\subsection{Traditional Einsums}\label{sec:background:einsums}
An Einsum defines a computation on a set of tensors over a set of points, called the \emph{iteration space}~\cite{itspace, teaal, edge}.
For example, we describe matrix-vector multiplication with the Einsum:
\begin{align}
\scalebox{1}{$
Z_{m} = A_{k, m} \times B_{k}
$}
\end{align}
This equation says, for each point $(k, m)$ in the iteration space, multiply the values of $A_{k, m}$ and $B_{k}$ and reduce the partial outputs across the 
$K$ rank to produce values of $Z_{m}$.
By convention, the index letter (e.g., $k$, $m$) is the lowercase form of the corresponding rank name (e.g., $K$, $M$).
Prior work~\cite{teaal, fusemax, edge} uses \emph{cascades}, or sequences of dependent Einsums, to describe a wider range of algorithms.

\subsection{Extended Einsums}\label{sec:background:edge}
Traditional Einsums can express standard tensor algebra. 
However, they cannot handle more complex computations.
The recently proposed Extended General Einsums notation (EDGE)~\cite{edge} extends Einsums to 
express graph algorithms. 
We now summarize the portions of EDGE we leverage.


EDGE separates computation into three \emph{actions}: map ($\bigwedge$), reduce ($\bigvee$), and populate ($\lll$).
Intermediate values produced between actions are called \emph{temporaries}.
The \emph{map} action is performed first and applies a computation to combine operands from the input tensors, producing the map temporaries.
The \emph{reduce} action then aggregates the map temporaries to produce the reduce temporaries.
Finally, the \emph{populate} action consumes the 
reduce temporaries and writes values to the output tensor.

Each action can be paired with user-defined \emph{compute} and \emph{coordinate operators},\footnote{The original EDGE paper used the term ``merge operators" for map and reduce; later work standardized the terminology as ``coordinate operators".}
enabling EDGE to express a wide range of algorithms.
The \emph{compute operator} specifies the operation applied to data values, while the \emph{coordinate operator} determines the regions of the iteration space where the computation is evaluated.
EDGE defines some common options for operators, but also supports user-defined custom operators.

\begin{figure}
     \centering
     \includegraphics[width=0.9\linewidth]{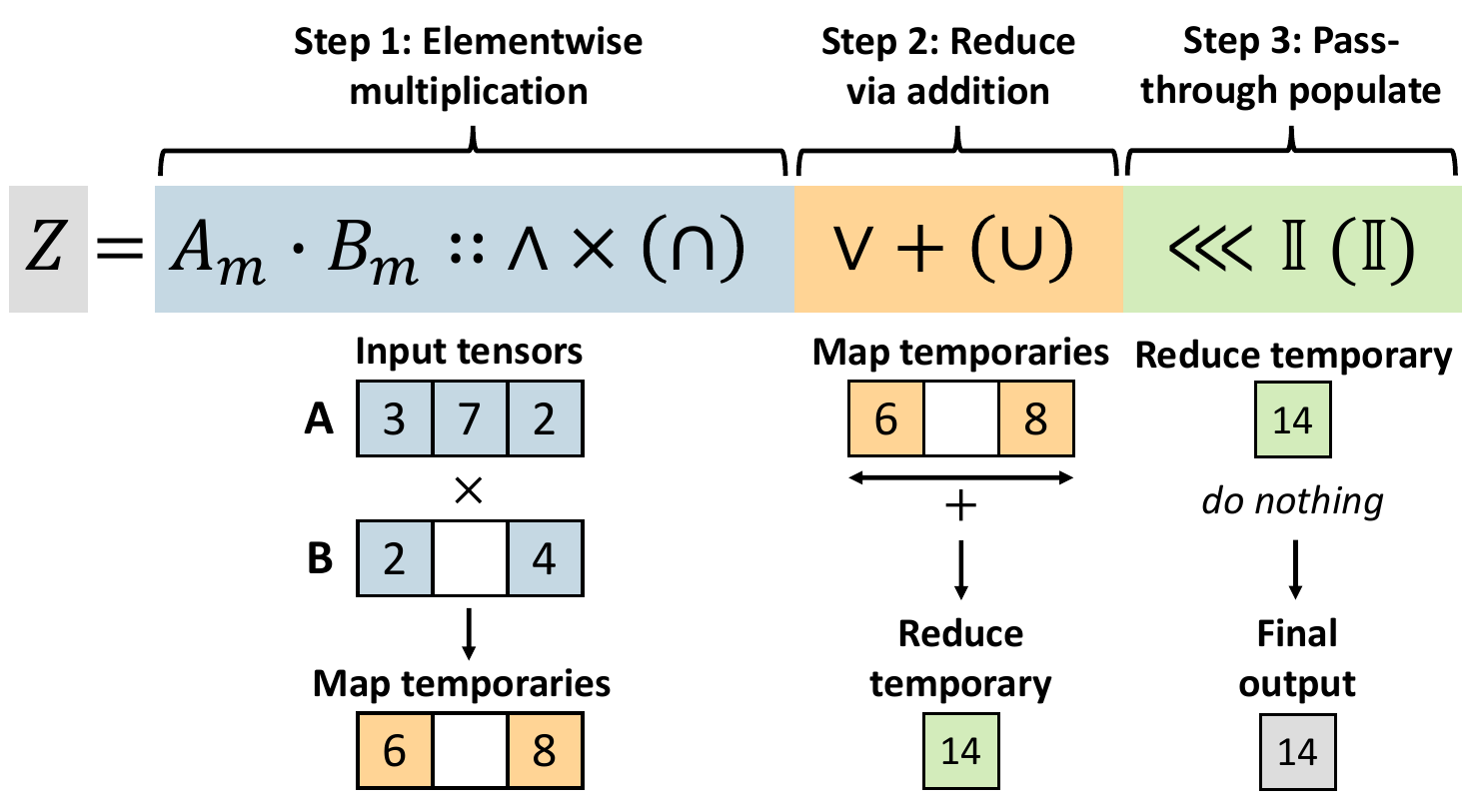}
     \caption{\textbf{Dot product Einsum and its stepwise breakdown. Example tensors are shown below each action to illustrate values consumed and  operations performed at each step. }
     }
     \label{fig:dot}
\end{figure}


Figure~\ref{fig:dot} shows the extended Einsum for the dot product between two input tensors $A_m$ and $B_m$.
The first step of a dot product is the elementwise multiplication of $A_m$ and $B_m$, highlighted in blue in Figure \ref{fig:dot}.
Accordingly, the map compute operator is ``$\times$'', indicating that values from the input tensors are combined via multiplication.
In sparse tensor algebra, multiplication is performed only at coordinates where both tensors contain nonzero values. This behavior is captured by the map coordinate operator, intersection ($\cap$), which defines the region of the iteration space over which the multiplication is evaluated.
After performing the elementwise multiplication, the next step is to sum the products (shown in orange in Figure~\ref{fig:dot}).
The reduce compute operator is a binary operator that combines a map temporary with the current reduce temporary.
In this case, we use addition.
If no current reduce temporary exists, the map temporary is copied into the reduce temporary.
The typical reduce coordinate operator that pairs with the ``$+$'' compute operator is
the union operator ($\cup$), indicating that the reduction is performed when either the map temporary or the current reduce temporary 
is nonzero.

After completing the map and reduce actions, the dot product result is already available. 
We therefore use the pass-through operator 
($\mathbbm{1}$) ~\cite{edge} as both the populate compute operator (i.e., performing no computation) and the populate coordinate operator (i.e., evaluating at all points), as shown in green in Figure \ref{fig:dot}.
When both the compute and coordinate operators of an action are 
pass-through, we 
will omit them from the Einsum notation (the populate action in Figure \ref{fig:dot} is shown only for completeness).
We also provide additional examples of populate coordinate operators in 
Appendix~\ref{app:ostar}.


\begin{figure}
     \centering
     \includegraphics[width=0.9\linewidth]{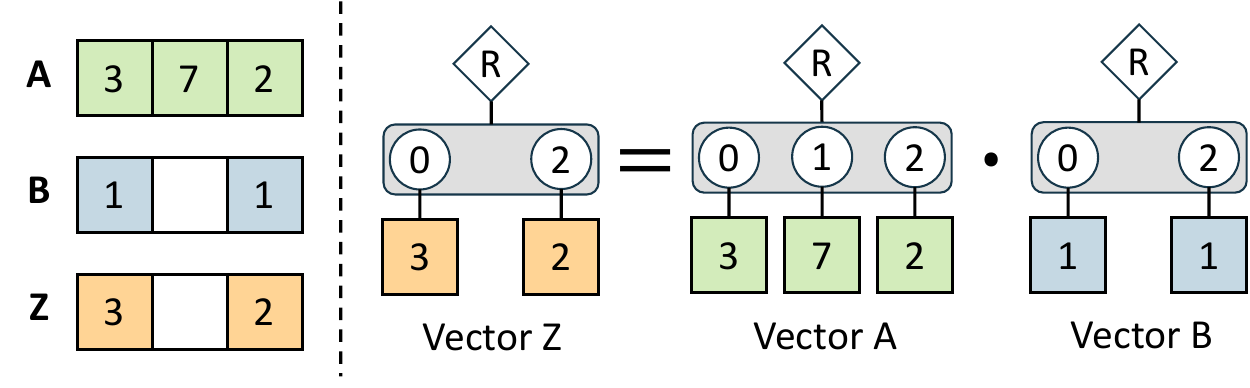}
     \caption{\textbf{Tensor and fibertree notation for illustrating take left ($\leftarrow$) and take right ($\rightarrow$) operators shown in Einsum~\ref{eq:elementmul_coord}.}
     }
     \label{fig:takeop}
\end{figure}

\para{Additional Common Operators}
In this work (specifically Section~\ref{sec:cascade}), we also use two other common operators: \emph{take-left} ($\leftarrow$) and \emph{take-right} ($\rightarrow$) ~\cite{edge}.
When these operators are used as coordinate operators, they select points where the corresponding input is non-empty.
When they are used as compute operators, they copy the corresponding input into the 
action's output. For instance, Figure \ref{fig:takeop} shows the tensor and fibertree notations for the following Einsum:
\begin{align}\label{eq:elementmul_coord}
\scalebox{1}{$
Z_{m} = A_{m} \cdot B_{m} :: \bigwedge \leftarrow (\rightarrow)
$}
\end{align}
This Einsum says that we output
the value from $A_m$ into $Z_m$ whenever the corresponding point in $B_m$ is non-empty.

When there is only one input tensor, we use take-left as the map coordinate operator to specify that the map compute operator should act on non-empty input values.
For example, the Einsum below copies all non-empty points in $A_m$ to $Z_m$:
\begin{align}\label{eq:copy}
\scalebox{1}{$
Z_{m} = A_{m} :: \bigwedge \mathbbm{1} (\leftarrow)
$}
\end{align}
Because there is only a single input tensor, no map compute operator is needed to combine values from multiple tensors; therefore, we use the pass-through operator as the map compute operator.

Finally, for the reduce action, the \emph{left} operator is the current reduce temporary and the \emph{right} operator is the new map temporary.
For example, the following Einsum sums together only the non-empty elements of $A_m$:
\begin{align}\label{eq:reduce}
\scalebox{1}{$
Z = A_{m} :: \bigwedge \mathbbm{1} (\leftarrow) \bigvee + (\rightarrow)
$}
\end{align}

\para{Iterative Ranks}
EDGE Einsums also support expressing loop-carried dependencies via iterative ranks.
Let $S_0 = 0$.
We express the prefix sum over array $A$ (Algorithm \ref{alg:prefix-sum})
with the following Einsum, where $I$ is the iterative rank:
\begin{align}
\scalebox{1}{$
S_{i + 1} = S_i \cdot A_i \label{eq:iter:s} :: \bigwedge + (\cup)
$}
\end{align}
 \begin{algorithm}
    \caption{Prefix Sum}\label{alg:prefix-sum}
    \begin{algorithmic}[1] 
        \State \textbf{for} $i = 0$ \textbf{to} $I-1$
            \State \hspace{0.3cm} $S_{i+1}$ = $S_i$ + $A_i$
    \end{algorithmic}
    \end{algorithm}

\subsection{TeAAL~\cite{teaal} Separation of Concerns}
\label{sec:background:pyramid}

\begin{figure}[h]
     \centering
    \includegraphics[width=0.35\textwidth]{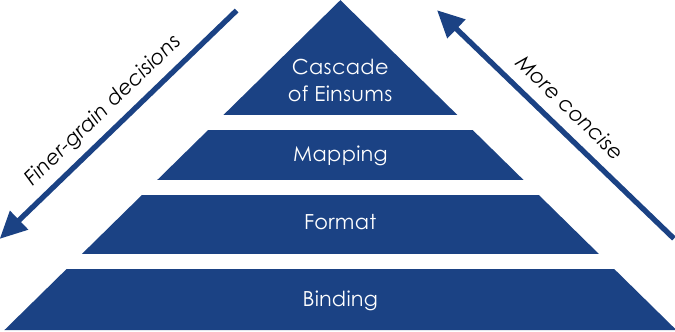}
     \caption{\textbf{The TeAAL separation of concerns for describing tensor algebra kernels ~\cite{teaal}.}}
     \label{fig:separation_of_concerns}
\end{figure}

Beyond the cascade of Einsums, this work uses the abstractions proposed by TeAAL
to describe the salient characteristics of the \ourscheme kernel.
The TeAAL abstractions provide a separation of concerns between these characteristics, which TeAAL organizes into the hierarchy shown in Figure~\ref{fig:separation_of_concerns}.
While the cascade of Einsums specifies what computation is required, the \emph{mapping}, \emph{format}, and \emph{binding} describe how it occurs.

\subsubsection{Mapping}
\label{sec:background:pyramid:mapping}

The mapping describes in what order the kernel visits points in the iteration space.
Attributes of mapping include partitioning (how the tensors and iteration space are split), loop order (the order points are visited), and spacetime schedule (which ranks are parallelized).

\subsubsection{Format}
\label{sec:background:pyramid:format}

The format describes the tensors' concrete representations.
TeAAL adopts a per-rank format, which allows each rank of fibers to use a different, custom layout.
Building on Section~\ref{sec:background:tensor}, TeAAL stores each fiber as 
coordinate and payload arrays, where the payload indicates the occupancy of the associated next-level fiber.
Thus, the format of a rank is defined by three parameters: 
\begin{itemize}
    \item \emph{(Un)compressed}: whether the coordinate and payload arrays are uncompressed (size proportional to shape) or compressed (size proportional to occupancy).
    \item \emph{cbits}: the bitwidth used for coordinates.
    \item \emph{pbits}: the bitwidth used for payloads.
\end{itemize}



\begin{figure}
\centering
\includegraphics[width=\linewidth]{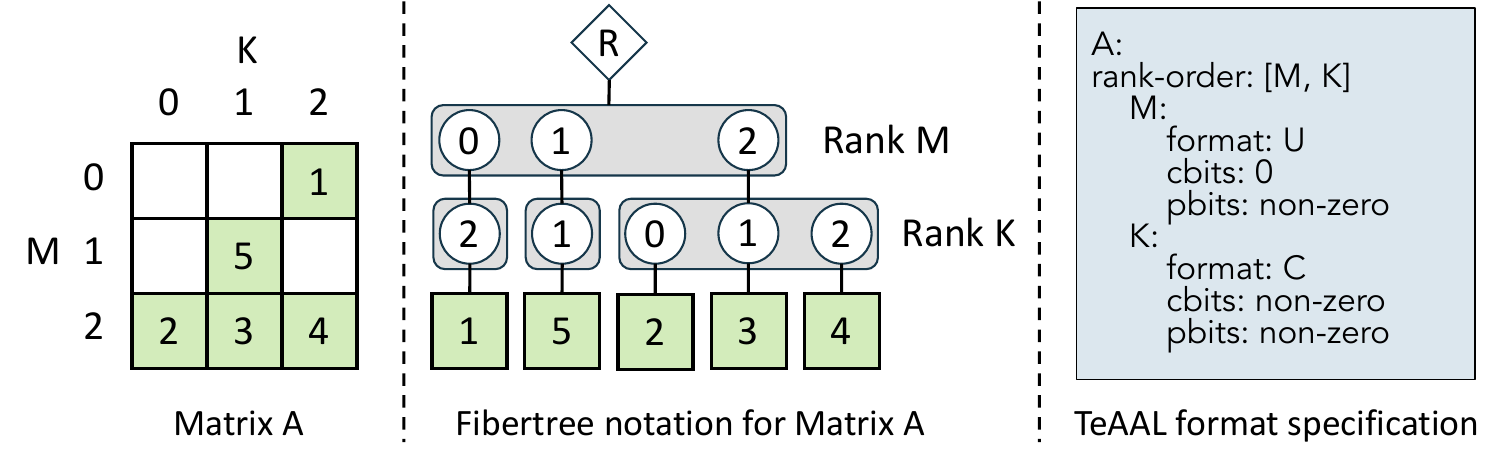} 
  \caption{\textbf{TeAAL format specification for a matrix $A$ encoded in the CSR format. 
  Each rank is annotated as either compressed (C) or uncompressed (U). 
  Note that in the uncompressed rank $M$, the coordinate bitwidth (\emph{cbits}) is zero because the coordinates are encoded implicitly by array position.
  }}
  \label{fig:background:format}
\end{figure}

Importantly, not all ranks need explicit coordinates and payloads.
For example, an uncompressed rank encodes coordinates implicitly through array position.
In this case, the cbits is set to zero.
Figure \ref{fig:background:format} shows an example TeAAL format specification for a matrix in the CSR format.

\subsubsection{Binding}
\label{sec:background:pyramid:binding}

The binding defines the spatial and temporal placement of tensors within the architecture, as well as where and when each point in the iteration space is executed.
Importantly, on a CPU, some aspects of binding (e.g., memory hierarchy placement) are transparent to software.
Therefore, in this work, our optimization of the binding focuses on how the mapped kernel is lowered into C++ and whether each 
component becomes data or instructions.

\section{Bottlenecks in Existing RTL Simulators}
\label{sec:motivation}

\label{sec:motivation:bottlenecks}

The high compilation costs and severe frontend bottlenecks in full-cycle RTL simulation are well known in the RTL simulation community.
Prior work has identified and analyzed these issues in Verilator, the state-of-the-art open-source simulator~\cite{Metro-Mpi,essent,essent-mag}.
These frontend bottlenecks are primarily caused by the large, statically generated C++ simulation code, 
and are exacerbated by branching.
ESSENT~\cite{essent,essent-mag} mitigates these issues by proposing an alternative approach that 
completely unrolls the RTL dataflow graph into straight-line code.
This design reduces branch overhead, improves instruction-cache (I-cache) prefetching, and enables more aggressive compiler optimizations.\footnote{While a key feature of ESSENT is its activity-aware \texttt{-O3} optimization, this work focuses on full-cycle, activity-oblivious simulation.
Accordingly, we consider ESSENT with \texttt{-O2}, which is the most performant activity-oblivious variant of ESSENT.}

We evaluate Verilator and ESSENT in terms of both simulation performance and compilation overhead across multiple RocketChip and SmallBOOM designs.
For simulation, we use the \texttt{dhrystone} benchmark.
All experiments are conducted on an AWS Graviton 4 machine and compiled with \texttt{clang -O3}.
The takeaway is that ESSENT is able to improve (but not eliminate) simulation-time bottlenecks, but does so at the cost of additional compilation time.

\begin{figure}[h]
    \centering
\includegraphics[width=0.95\linewidth]{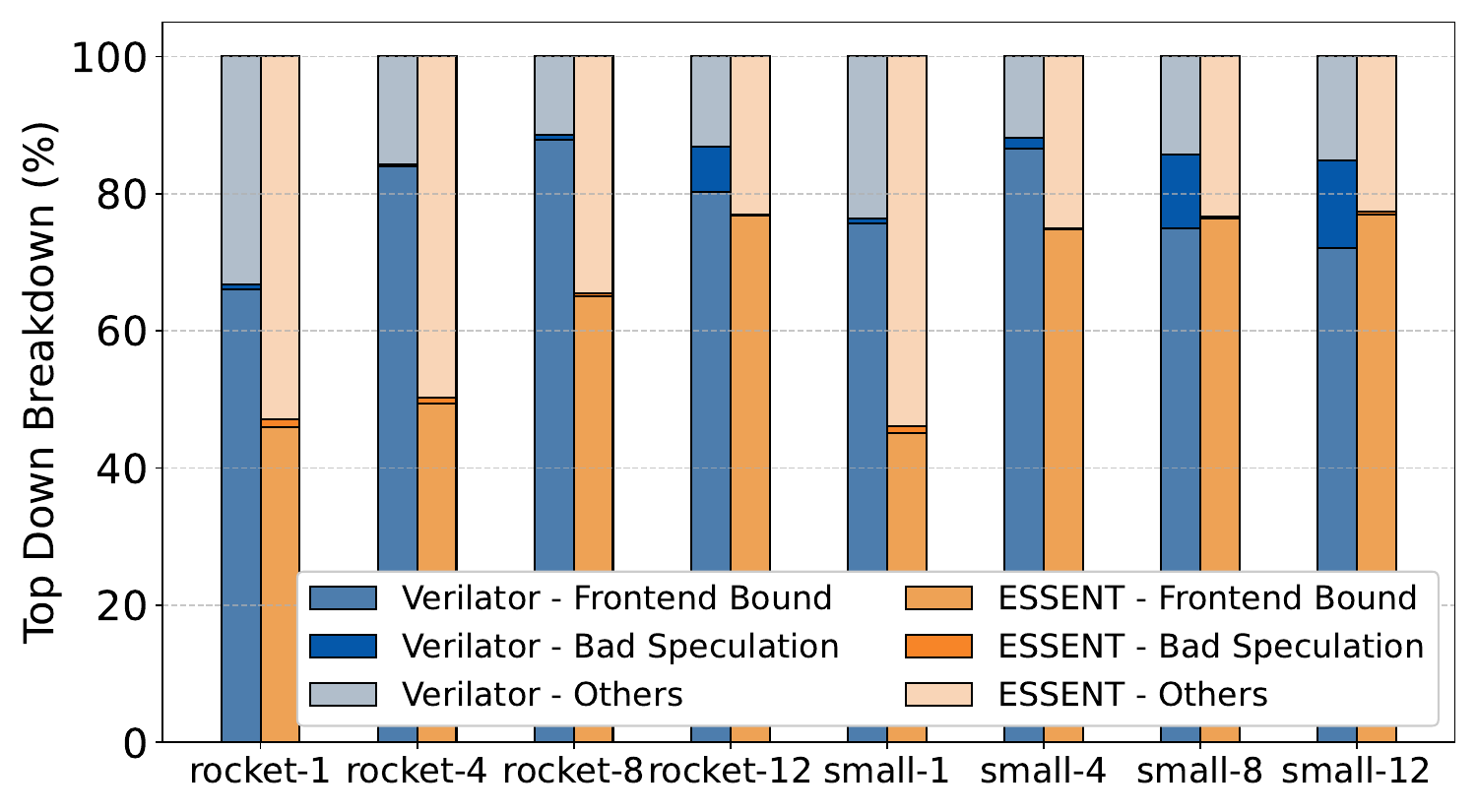} 
    \caption{
    \textbf{
    Top-down breakdown of Verilator and ESSENT for 1 to 12-core RocketChip and SmallBOOM simulations. ``Others” aggregates backend-bound and retiring categories.
    }}
    \label{fig:frontend}
\end{figure}

Figure~\ref{fig:frontend} shows the results of a 
top-down analysis~\cite{topdown} for Verilator and ESSENT during simulation.
ESSENT consistently exhibits a lower fraction of frontend-bound and bad-speculation slots than Verilator.
We further measure the L1 I-cache miss per kilo instructions (MPKI) for these designs.
Verilator incurs between 80 and 120 MPKI, while ESSENT reduces this to between 64 and 70 MPKI, indicating improved instruction locality but still substantial frontend pressure.


\begin{figure}[h]
    \centering
\includegraphics[width=0.9\linewidth]{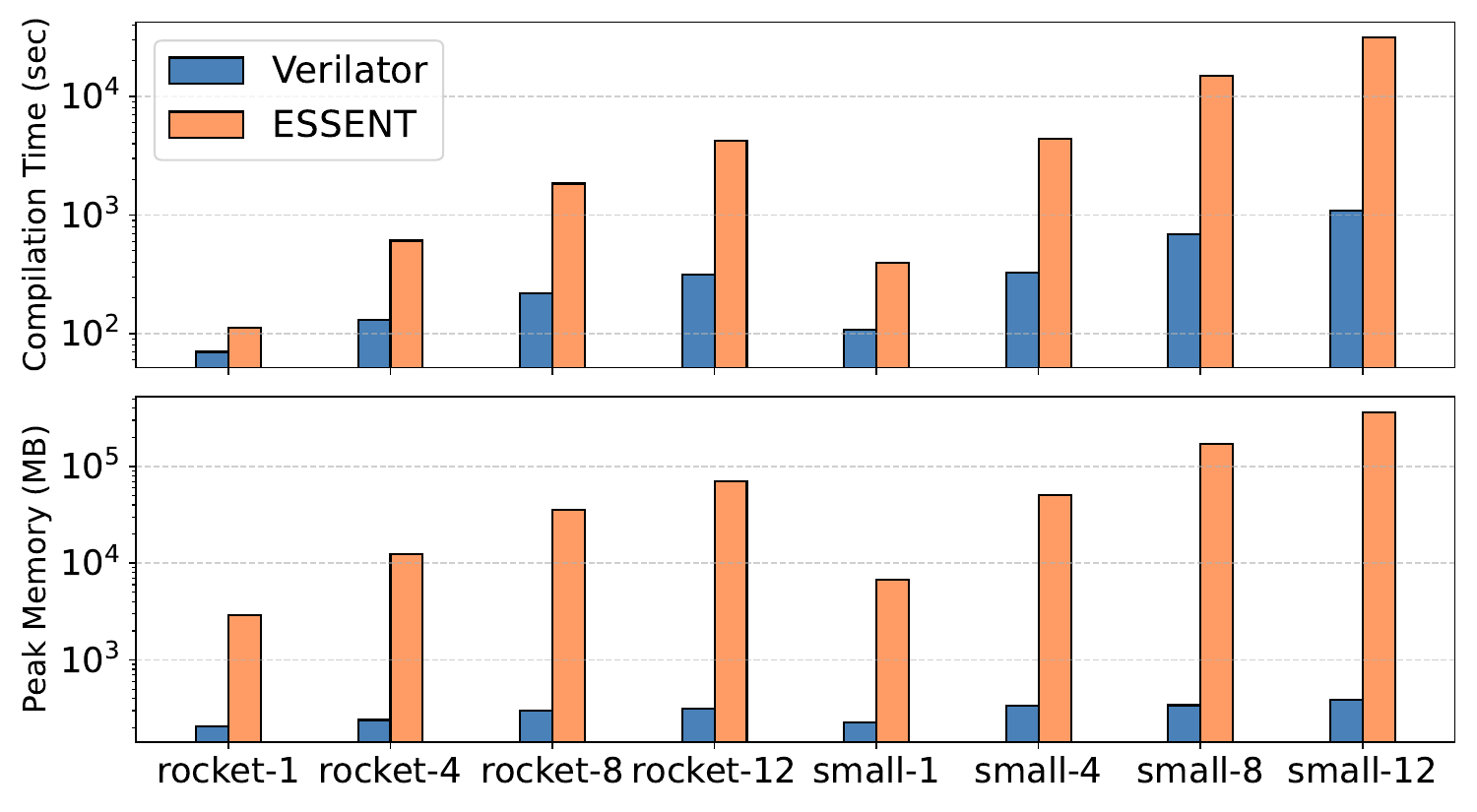} 
    \caption{
    \textbf{Compilation costs for Verilator and ESSENT when simulating 1 to 12-core RocketChips and SmallBOOMs.
    The top bar plot shows the compilation time, while the bottom one shows the peak compilation memory usage.
    }}
    \label{fig:stack}
\end{figure}
However, ESSENT’s approach increases code size and, consequently, compilation cost. Figure~\ref{fig:stack} reports the compilation time (top) and peak memory usage (bottom) for Verilator and ESSENT.
Note that the y-axes use a logarithmic scale.

Together, these results motivate a tensor algebra-based approach, which can represent computation using a more compact binary, addressing both frontend bottlenecks and compilation overhead.
In Section \ref{sec:cascade}, we show how to formulate RTL simulation as a tensor algebra problem.

\label{sec:motivation:tensor-algebra}

\section{RTL Simulation as a Cascade of Einsums}
\label{sec:cascade}

This section develops a
cascade of Einsums 
that captures the behavior of \emph{arbitrary} synchronous RTL designs as a tensor algebra kernel.
For simplicity, we assume a single clock domain (multi-clock support is described in Section~\ref{sub:system-integration}).
In Section~\ref{sec:cascade:independent}, we start by writing an Einsum for one layer of dataflow graph operations (a set of operations without dependencies).
Then, in Section~\ref{sec:cascade:arbitrary-graphs}, we extend our representation to arbitrary dataflow graphs.

\subsection{Evaluating Single-Layer Operations}
\label{sec:cascade:independent}

\begin{figure}[t]
    \centering
    \begin{subfigure}[t]{0.45\linewidth}
    \centering
\includegraphics[width=0.9\linewidth]{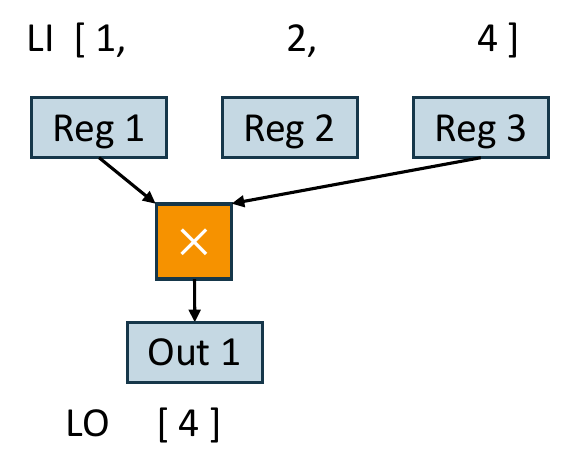} 
    \caption{Single multiply}
    \label{fig:DFG-a}
    \end{subfigure}
    \hfill
    \begin{subfigure}[t]{0.45\linewidth}
    \centering
    \includegraphics[width=0.9\linewidth]{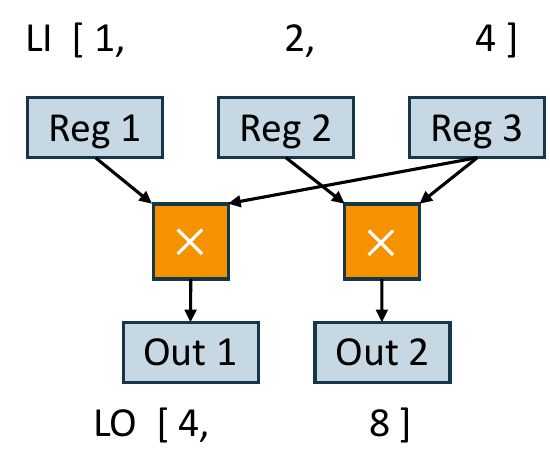} 
    \caption{Two multiplies}
    \label{fig:DFG-b}
    \end{subfigure}
    \caption{
    \textbf{Example dataflow graphs with register inputs 1, 2, and 4. The inputs form the input tensor $LI$, with rank $R$.
    }}
    \label{fig:DFG}
\end{figure}

\begin{figure}[t]
    \centering

    \begin{subfigure}[t]{0.9\linewidth}
        \centering
        \includegraphics[width=\linewidth]{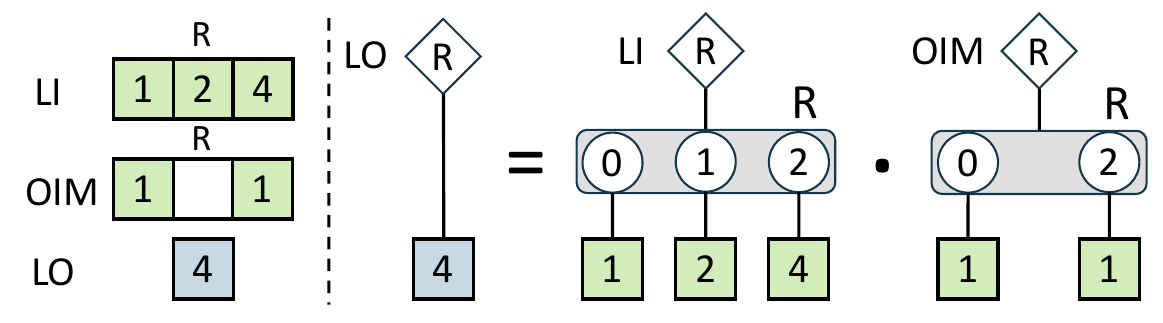}
        \caption{
        Tensors and fibertrees for Figure~\ref{fig:DFG-a}. Nonempty payloads in $OIM$ identify operands in $LI$, which are multiplied, and the resulting product (4 in this example) is stored in $LO$.
        }
        \label{fig:Rrank}
    \end{subfigure}


    \begin{subfigure}[t]{0.9\linewidth}
        \centering
        \includegraphics[width=\linewidth]{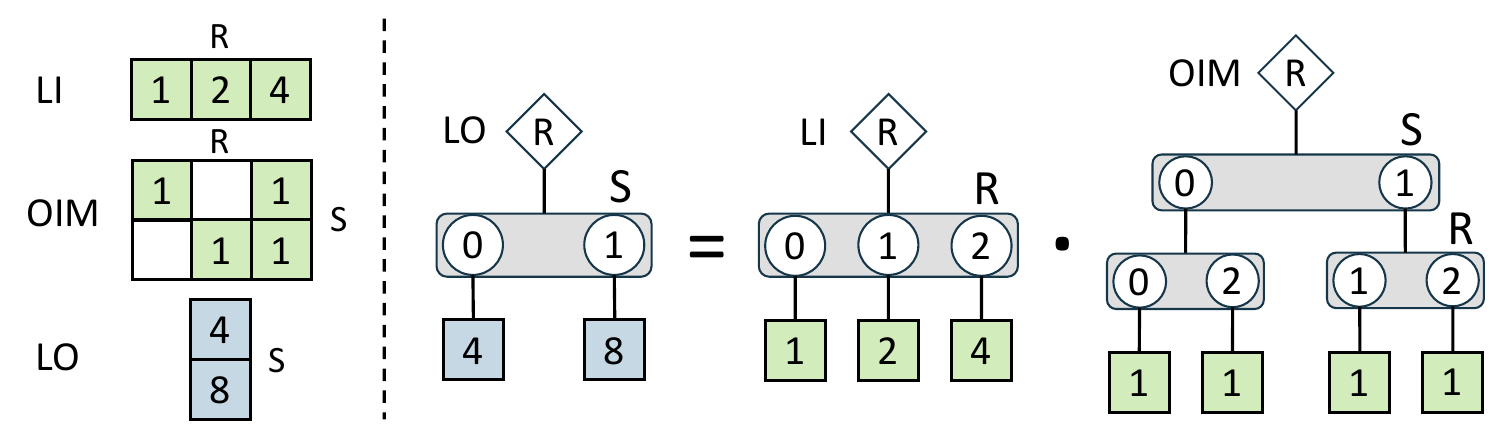}
        \caption{
        Tensors and fibertrees for Figure~\ref{fig:DFG-b}. The $S$ rank is added to $OIM$. $LO$ also gains an $S$ rank to store two outputs.
        }
        \label{fig:Srank}
    \end{subfigure}

    \caption{
    \textbf{ The $LI$, $OIM$, and $LO$ tensors and fibertree representations for dataflow graphs in Figure~\ref{fig:DFG}.}
    }
    \label{fig:rank_examples}

\end{figure}

\para{Evaluating a Single Operation} 
We begin by expressing an RTL dataflow graph made up of a single multiply operation as a cascade of Einsums.
Figure \ref{fig:DFG-a} shows the corresponding dataflow graph.
First, we define a tensor $\layerinput$ (layer input), which specifies the register inputs to the dataflow graph, denoted by rank $R$.
Next, we introduce a binary mask tensor $OIM$ (operation input mask) that identifies which elements of $LI$ serve as operation inputs.
Finally, we define a tensor 
$LO$ (layer output) that stores the resulting output.
Figure~\ref{fig:Rrank} illustrates these tensors for the dataflow graph in Figure~\ref{fig:DFG-a}, along with the corresponding 
fibertrees.
We can describe the multiply with the following cascade:
\begin{align}\label{eq:step1}
\scalebox{1}{$
\layeroutput = \layerinput_r \cdot \operationinputmask_r :: 
\bigwedge \leftarrow (\rightarrow) \bigvee \times (\rightarrow)
$}
\end{align}
We read this cascade as follows: after the map action selects operands from $LI$ based on coordinates with non-empty payload in $OIM$, we
reduce them with multiplication ``$\times$''.



\para{Evaluating Multiple Operations}
We now extend the cascade from a single multiply operation to multiple multiplies.
We introduce an additional rank, $S$, into $\layeroutput$ and $\operationinputmask$
(indicating $S$ outputs).
This extended cascade is:
\begin{align}\label{eq:step2}
\scalebox{1}{$
\layeroutput_s = \layerinput_r \cdot \operationinputmask_{r,s} :: 
\bigwedge \leftarrow (\rightarrow) \bigvee \times (\rightarrow)
$}
\end{align}

Figure~\ref{fig:DFG-b} shows a dataflow graph containing two multiply operations.
Figure~\ref{fig:Srank} presents the corresponding tensors and their fibertree representations.
As shown, the $S$ rank is added to $OIM$ to encode two multiplies, making it a 
matrix.
Similarly, $\layeroutput$ gains an $S$ rank of shape 2, producing two outputs, one for each multiply.


\para{Evaluating Multiple Operation Types}
We now generalize our cascade to support $N$ distinct operation types, extending several parts of the Einsum. 
First, we add an $N$ rank to the $OIM$, which specifies the operation type for each operation ($s$ coordinate).
We categorize the operation types into 3 classes: \emph{reducible}, \emph{unary}, and \emph{select operations} and develop their Einsum formulations one by one, below.
    
\textit{Reducible operations} are those in which map temporaries can be combined pairwise using a reduce compute operator.
After the map action selects operands from $\layerinput$, the reduce compute operator aggregates them to produce the final output.
To support multiple reducible operations in the same cascade,  
we replace the previously fixed operator (e.g., the multiply operator “$\times$” in Einsum~\ref{eq:step2}) with the custom operator $op\_r[n]$.
This operator implements a case statement that performs different computation depending on the value of $n$, as shown in Algorithm~\ref{alg:opn}.
When $n$ corresponds to a non-reducible operation, the operator copies the map temporary into the reduce temporary.
The resulting cascade is:
\begin{align}
\scalebox{1}{$
\layeroutput_{s} = \layerinput_r \cdot \operationinputmask_{n, r, s} :: 
\bigwedge \leftarrow (\rightarrow) \bigvee\, op\_r[n] (\rightarrow)
$}
\end{align}
\begin{algorithm}
    \caption{Custom operator $op\_r[n]$ for operation type $n$}
    \label{alg:opn}
    \begin{algorithmic}[1]
        \small
        \State \textbf{function} $op\_r[n](prev\_out, map\_tmp)$
        \State \hspace{0.6cm} \textbf{switch} $(n)$
        \State \hspace{1.2cm} \textbf{case} $0$: 
        \Comment{{\footnotesize\;multiply}}
        
        \State \hspace{1.5cm} $cur\_out \gets prev\_out \times map\_tmp$
        \State \hspace{1cm}  \textbf{...}
        \State \hspace{0.6cm} \Return $cur\_out$
        \State \textbf{end function}
    \end{algorithmic}
\end{algorithm}

Unfortunately, not all operations are commutative (e.g., subtraction), which means that the order that map temporaries are passed to the reduce action is critical for correctness.
We encode this ordering using a new rank, denoted as $O$.
The $O$ rank defines the order in which operands are passed to the reduce action.
That is, the $r$ coordinate whose $o$ coordinate equals 0 is passed to reduce first, the $r$ coordinate whose $o$ coordinate equals 1 is passed to reduce second, and so on. 
In effect, the $O$ rank permutes the order that $r$ coordinates are visited.

The above usage of the $O$ rank assumes that $o$ coordinates are visited in the iteration space in coordinate-ascending order.
In traditional Einsum notation, traversal order is intentionally left undefined to allow maximum flexibility during scheduling, and ordering constraints are described solely by the mapping and binding.
However, as we extend the Einsum formalism to support a broader class of operators, ordering constraints at the Einsum level become necessary.
In this work, we (informally) impose this ordering constraint on the $O$ rank to ensure correctness in the face of non-commutative operators.
We leave a more formal treatment of traversal order constraints for extended Einsums to future work.


\begin{align}\label{eq:custom_reduce}
\scalebox{1}{$
\layeroutput_{s} = \layerinput_r \cdot \operationinputmask_{n, o, r, s} :: 
\bigwedge \leftarrow (\rightarrow) \bigvee\, op\_r[n] (\rightarrow)
$}
\end{align}

\textit{Unary operations}, in contrast, take a single input (e.g., negation). 
Because the Einsum only performs the reduce compute operator when multiple map temporaries try to update the same reduce temporary, unary operations cannot be performed as a part of $op\_r[n]$.
Instead, we would like to handle unary operations via the map compute operator. 
To allow for a second map compute operator, we first split Einsum~\ref{eq:custom_reduce} into a cascade of two Einsums, as follows:
\begin{align}\label{eq:split_einsum1}
\scalebox{1}{$
OI_{n, o, r, s} = LI_{r} \cdot OIM_{n, o, r, s} ::  \bigwedge \leftarrow (\rightarrow)
$}
\end{align}
\begin{align}\label{eq:split_einsum2}
\scalebox{1}{$
\layeroutput_{n, s} = OI_{n, o, r, s} ::  \bigwedge \mathbbm{1} (\leftarrow) \bigvee op\_r[n] (\rightarrow)
$}
\end{align}
Einsum~\ref{eq:split_einsum1} uses $OIM$ to select the elements of the input $LI$ and place them in a newly named operation input tensor $OI$.
Notice that $OI$ holds the same values as the map temporary from Einsum~\ref{eq:custom_reduce}.
Then, Einsum~\ref{eq:split_einsum2} first uses the map action to retrieve the non-empty values of $OI$ (see Section~\ref{sec:background:edge}) and reduces them together with $op\_r[n]$.

We can now replace the pass-through map compute operator in Einsum \ref{eq:split_einsum2} with the custom operator $op\_u[n]$, implemented similarly to $op\_r[n]$, 
to support unary operations:

\begin{align}\label{eq:step4c}
\scalebox{1}{$
\layeroutput_{n, s} = OI_{n, o, r, s} ::  \bigwedge op\_u[n] (\leftarrow) \bigvee op\_r[n] (\rightarrow)
$}
\end{align}

Finally, we support \textit{select operations}, such as conditional select (mux).
Given a selector and two candidate inputs, the operator chooses one input. Because this behavior is neither reducible (it does not combine operands pairwise) nor unary, it cannot be expressed using either the map or reduce compute operators. 
Instead, we use the populate coordinate operator, which acts on an entire $O$-fiber of the reduce temporary. 
In this case, we use the custom operator $op\_s[n]$ to express that select operations first collect all inputs before deciding which one to keep in the output.

Because Einsum~\ref{eq:step4c} requires reducing over the $O$ rank and the select operations require preserving the $O$ rank in the reduce temporary, we add a third Einsum to our cascade:
\begin{align}\label{eq:step5b}
\scalebox{1}{$
\layeroutput\_sel_{n, o*, r, s} = OI_{n, o, r, s} ::  \bigwedge \mathbbm{1} (\leftarrow) \lll \mathbbm{1} (op\_s[n])
$}
\end{align}
Notice that $LO\_sel$, which stores the output of select operations, has a special rank variable expression $o*$.
We present a detailed explanation of the populate coordinate operator and rank variable expression $o*$ in
Appendix~\ref{app:ostar}. 
Together, Einsums \ref{eq:split_einsum1}, \ref{eq:step4c}, and \ref{eq:step5b}
form the cascade for executing a single layer of the dataflow graph.

\subsection{Evaluating Arbitrary Dataflow Graphs}
\label{sec:cascade:arbitrary-graphs}

\begin{figure}
    \centering
    \begin{subfigure}[t]{0.45\linewidth}
    \centering
    \includegraphics[width=0.77\linewidth]{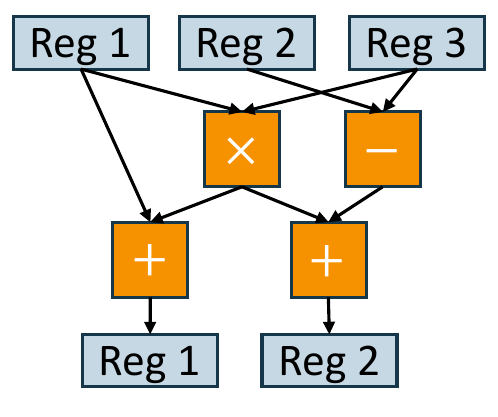} 
    \caption{Original dataflow graph}
    \label{fig:df_preprocess-a}
    \end{subfigure}
    \hfill
    \begin{subfigure}[t]{0.45\linewidth}
    \centering
    \includegraphics[width=0.8\linewidth]{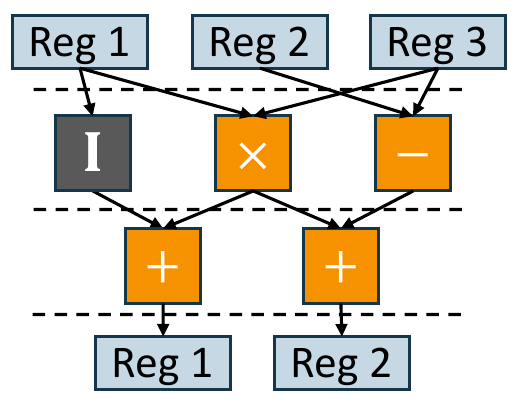} 
    \caption{Preprocessed graph}
    \label{fig:df_preprocess-b}
    \end{subfigure}
    \caption{\textbf{The original dataflow graph (a) is sliced into layers based on data dependencies, and identity operators (gray squares) are inserted to break cross-layer dependencies (b).
}}
\label{fig:df_preprocess}
\end{figure}

To extend our cascade from a single layer of operations to an arbitrary dataflow graph, we first perform a two-step preprocessing of the dataflow graph, as shown in Figure~\ref{fig:df_preprocess}.
First, we perform \emph{levelization}~\cite{essent}, slicing the graph into layers so that each operation depends only on outputs from layers above it.
Next, we remove cross-layer dependencies by propagating values forward, ensuring that each layer $i{+}1$ depends only on the outputs of layer $i$.
We refer to these value-propagation operations as \emph{identity operations}.

Finally, to iterate over the full multi-layer dataflow graph, we add an iterative rank $I$
to all tensors. 
Figure \ref{fig:tensor} shows the fibertree for OIM with all of its ranks.
We also introduce an Einsum to propagate the outputs of layer $i$ to the inputs of layer $i{+}1$ ($LI$). Specifically, when the $n$ coordinate corresponds to reducible or unary operations, the Einsum copies $LO$; when $n$ represents select operations, it copies $LO\_sel$.
The complete cascade is shown in Cascade~\ref{box:einsum}.
\newtcolorbox[auto counter]{einsumbox}[2][]{%
  colback=white,
  colframe=black,
  boxrule=0.5pt,
  sharp corners,
  arc=0pt,
  label={#1}
}

\begin{figure}
\centering
\small
\begin{einsumbox}[box:einsum]{}

\[
\begin{aligned}
OI_{i, n, o, r, s} &= \layerinput_{i, r} \cdot \operationinputmask_{i, n, o, r, s} :: \textstyle\bigwedge \leftarrow (\rightarrow) \\
\layeroutput_{i, n, s} &= OI_{i, n, o, r, s} :: \textstyle\bigwedge op\_u[n] (\leftarrow) \textstyle\bigvee\ op\_r[n](\rightarrow) \\
\layeroutput\_sel_{i, n, o*, r, s} &= OI_{i, n, o, r, s} :: \textstyle\bigwedge \mathbbm{1} (\leftarrow) \lll \mathbbm{1}(op\_s[n])
\end{aligned}
\]

\[
\scalebox{1}{$
\layerinput_{i+1, s} =
\left\{
\begin{array}{l}
\layeroutput_{i, n, s} :: \bigwedge \mathbbm{1} (\leftarrow) \bigvee\, ANY(\rightarrow), n \notin n\_sel \\
\layeroutput\_sel_{i, n, o, r, s} :: \bigwedge \mathbbm{1} (\leftarrow) \bigvee\, ANY(\rightarrow), n \in n\_sel
\end{array}
\right.
$}
\]

\[
\scalebox{1}{$
\diamond : i \equiv I
$}
\]
\end{einsumbox}
\smallskip
\noindent\small
\textbf{Cascade 1: \ourscheme Einsum Cascade.
}
\end{figure}

Algorithm~\ref{alg:RU} provides pseudocode illustrating the computation performed by this cascade, assuming the loop order is $[I, S, N, O, R]$.
Note that fibers of the $N$ and $R$ ranks of $OIM$ are one-hot: each operation ($s$ coordinate) has a single type ($n$ coordinate) and each operand order index ($o$ coordinate) 
has
one input operand ($r$ coordinate).
Since $s$ coordinates are unique across operations within a layer, the $LO$ and $LO\_sel$ tensors in Cascade~\ref{box:einsum} do not overlap, allowing us to use a single \texttt{LO} array to collect all layer outputs
in Algorithm \ref{alg:RU}.
\begin{algorithm}
    \caption{Example kernel for RTeAAL Sim cascade}
    \label{alg:RU}
    \begin{algorithmic}[1]
        \small
        \State \textbf{for} $i = 0$ \textbf{to} $I-1$ \textbf{do} 
        \Comment{{\footnotesize Rank $I$}}
        \State \hspace{0.3cm} $\_, i\_p \gets OIM\_i[i]$
        \State \hspace{0.3cm} \textbf{for} $s = 0$ \textbf{to} $i\_p-1$ \textbf{do} \Comment{{\footnotesize Rank $S$}}
        \State \hspace{0.6cm} $n\_c, n\_p = OIM\_n$.next()  \Comment{{\footnotesize Rank $N$}}
        \State \hspace{0.6cm} \textbf{for} $o = 0$ \textbf{to} $n\_p-1$ \textbf{do} \Comment{{\footnotesize Rank $O$}}
        \State \hspace{0.9cm}  $r\_c, \_ = OIM\_r$.next() \Comment{{\footnotesize Rank $R$}}
        \State \hspace{0.9cm} $sel\_inputs[o] = LI[r\_c]$
        \State \hspace{0.9cm} $map\_tmp = op\_u[n\_c](LI[r\_c])$ 
        \State \hspace{0.9cm} $LO[s] = op\_r[n\_c](LO[s], map\_tmp)$ 
        \State \hspace{0.6cm} \textbf{if} $n\_c$ in $n\_sel$
        \Comment{{\footnotesize if $n\_c$ corresponds to select operations}}
        \State \hspace{0.9cm} $LO[s] = op\_s[n\_c](sel\_inputs)$
        \State \hspace{0.3cm} \textbf{for} $s = 0$ \textbf{to} $i\_p-1$ \textbf{do} \Comment{{\footnotesize Rank $S$}}
        \State \hspace{0.6cm} 
        $s\_c, \_ = OIM\_s$.next()
        \State \hspace{0.6cm} 
        $LI[s\_c] = LO[s]$
        \Comment{{\footnotesize Write $LO$ back to $LI$}}
    \end{algorithmic}
    \parbox{\linewidth}{\footnotesize\emph{Note:} The \texttt{next()} method is used to traverse coordinates (denoted by $\_c$) and payloads (denoted by $\_p$) within the fibers of a rank.}
\end{algorithm}

\subsection{Identity Operator Elision} 
\label{sec:key-opts:identity}

In Section~\ref{sec:cascade:arbitrary-graphs}, we introduced identity operators for cascade construction. 
However, real designs require far more identity operators than all other operations combined (Table~\ref{fig:key-opts:identity-ops}).
Therefore, to approach the performance of existing simulators, we must eliminate the cost of explicitly copying values for identity operations.
An identity operation becomes redundant whenever it has the same source and destination coordinates. 
We modify the $\operationinputmask$ so that all identity operations use matching source ($R$) 
and destination ($S$) 
coordinates, allowing us to elide explicit references to these now-redundant operations. 

\begin{table}[h]
\centering
\caption{\textbf{Required identity operations for different RocketChip and SmallBOOM configurations.}
}
\label{fig:key-opts:identity-ops}
\footnotesize
\resizebox{\columnwidth}{!}{%
\begin{tabular}{p{3cm}cccc}
\toprule
 & Rocket-1c & Small-1c & Rocket-8c & Small-8c \\ 
\midrule
Effectual Operations & 60 K  & 94 K  & 139 K  & 281 K  \\ 
Identity Operations  & 414 K & 891 K & 957 K  & 2992 K \\ 
\bottomrule
\end{tabular}%
}

\end{table}

\section{Tensor Algebraic Optimizations for RTL Simulation}
\label{sec:optimization}

After deriving the Einsum representation for RTL simulation, we proceed down the TeAAL hierarchy to examine optimizations at the mapping, 
format, and binding levels.

\subsection{Format Optimization -- Tensor Compression}
\label{sec:format}

The fibertree abstraction (Section \ref{sec:background:tensor}) enables the design of custom compressed formats.
Beyond reducing data movement, compression also decreases the number of load instructions required.
Thus, compression provides benefit whether a kernel is memory-bound or dynamic instruction-bound. 

\begin{figure}[h]
    \centering
    \includegraphics[width=0.95\linewidth]{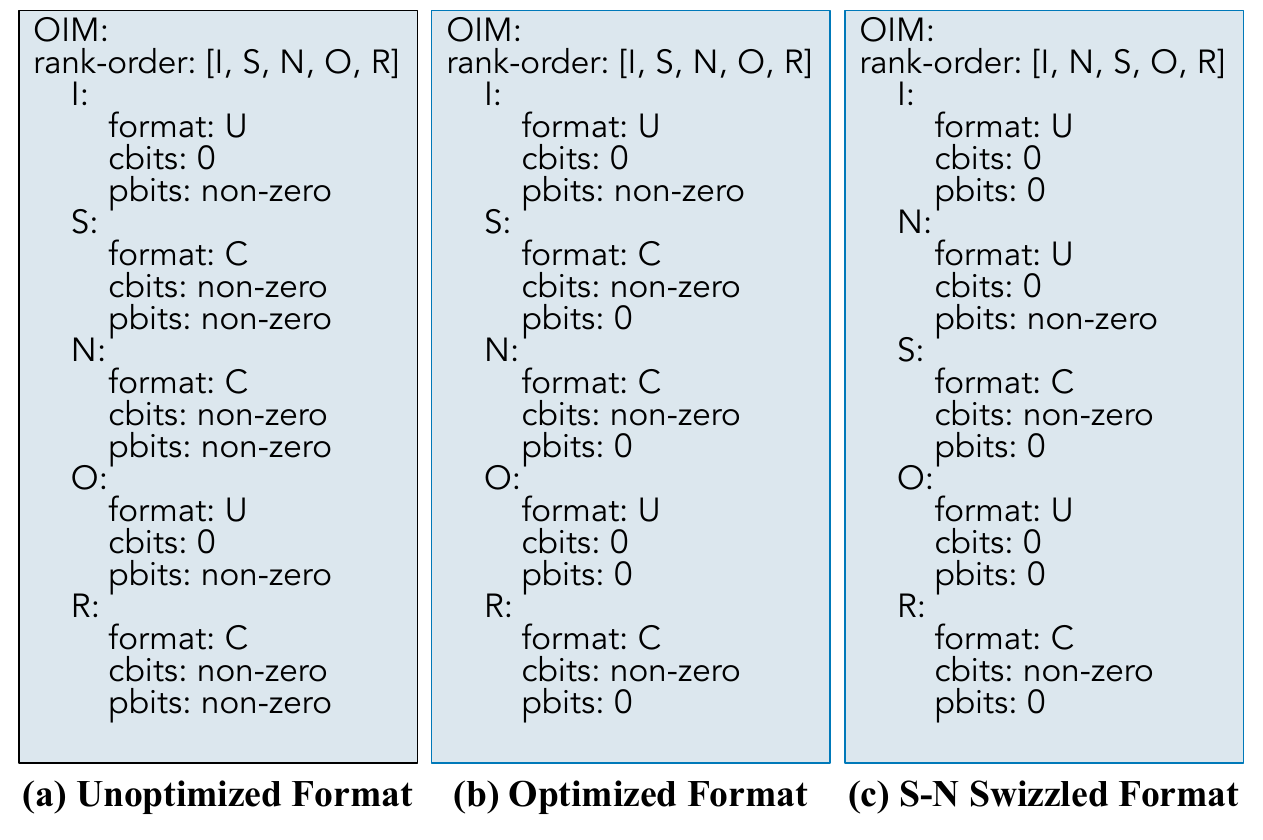}
    \caption{\textbf{Stepwise format optimization of the tensor $\operationinputmask$ toward a more compact and efficient representation.}
    }
    \label{fig:opt-space:format}
\end{figure}

\para{Lowering \operationinputmask{} to a Concrete Format}
In the \ourscheme cascade, $\operationinputmask$ 
has a density between $10^{-7}$ and $10^{-9}$,
making it a prime candidate for a custom sparse tensor format. 
Figure~\ref{fig:opt-space:format} describes the lowering of the $\operationinputmask$ fibertree 
onto a concrete format.
First, we recognize that fibers in ranks $I$ and $O$ are dense, so should be uncompressed ($U$); while fibers in ranks $S$, $N$, and $R$ are sparse, so should be compressed ($C$).
Figure~\ref{fig:opt-space:format}a lowers each rank onto its coordinate and payload arrays.
As noted in Section~\ref{sec:background:pyramid:format}, uncompressed ranks do not need explicit coordinates, so their $cbits$ are set to zero.
The bit width of each non-zero field is determined offline based on the maximum value for that coordinate or payload array.

\begin{figure}[h]
    \centering
    \includegraphics[width=\linewidth]{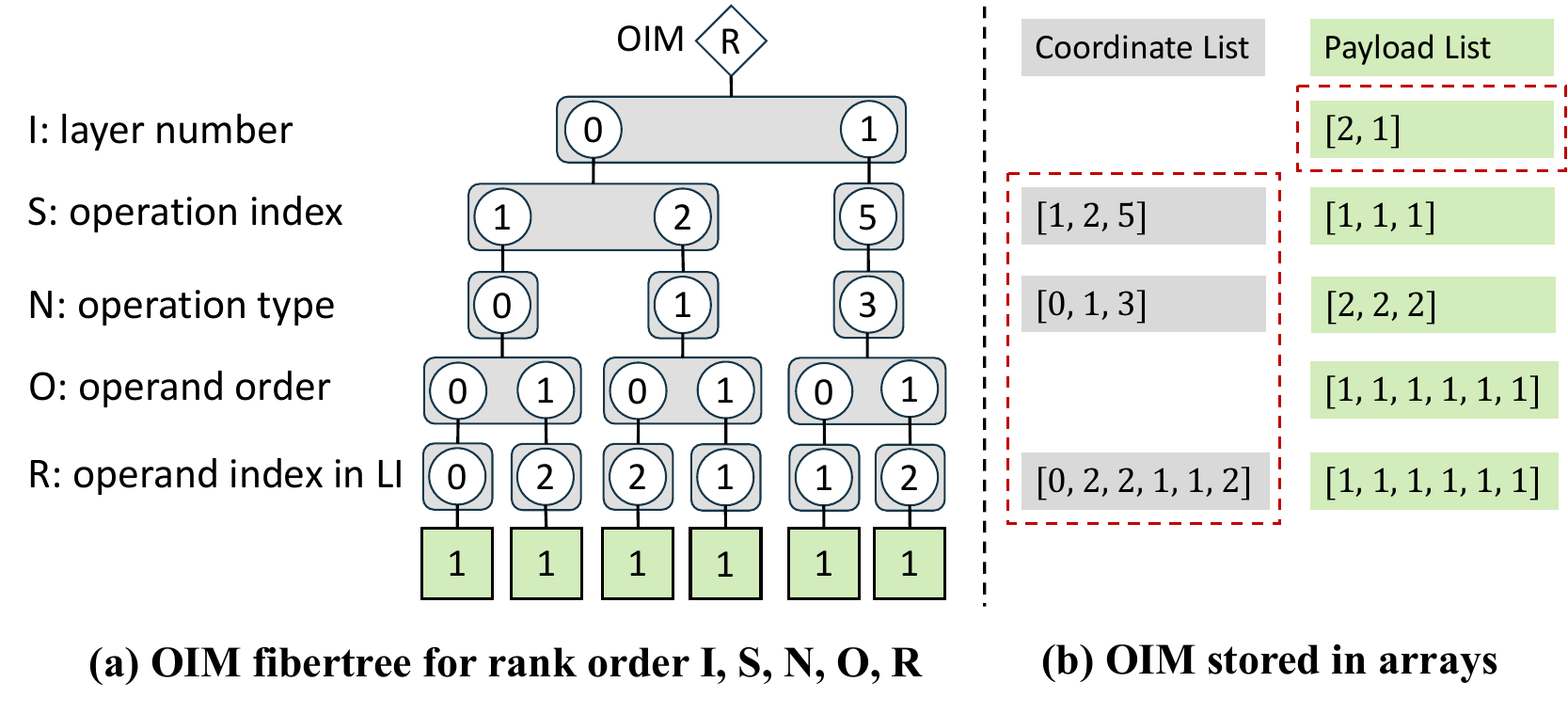}
    \caption{\textbf{The example $OIM$ (a) and its concrete representation (b). The unoptimized format (Figure~\ref{fig:opt-space:format}a) uses all arrays, while the optimized format (Figure~\ref{fig:opt-space:format}b) uses only the arrays highlighted with red dashed boxes.}}
    \label{fig:tensor}
\end{figure}

\para{Compressing the \operationinputmask{} Format}
We exploit the structure of the $OIM$ to further compress its representation.
As discussed in Section~\ref{sec:background:pyramid:format}, payloads encode the occupancy of the next-rank fiber.
For one-hot ranks ($N$ and $R$), fiber occupancy is always one, making payloads in the ranks above ($S$ and $O$) redundant. 
Moreover, the operation type ($N$) determines the number of input operands (the occupancy of the $O$ rank fiber), so payloads in the $N$ rank are also redundant.
Finally, because $OIM$ is a mask, the presence or absence of a coordinate already encodes whether the value is $1$ or $0$, rendering payloads in the $R$ rank unnecessary. 
Eliminating these payload arrays (i.e., setting their $pbits$ to zero) yields the optimized format shown in Figure~\ref{fig:opt-space:format}b.

To illustrate the effect of format compression, we present a toy example showing how the storage format changes before and after compression. Figure~\ref{fig:tensor}a shows the fibertree of $OIM$ with rank order 
$[I,S,N,O,R]$.
Figure~\ref{fig:tensor}b shows all arrays included in the unoptimized format from Figure~\ref{fig:opt-space:format}a; the tensors highlighted with red dashed boxes correspond to the optimized format in Figure~\ref{fig:opt-space:format}b.


In Section~\ref{sec:unrolling} (starting at NU), to optimize our kernel, we swap (swizzle~\cite{itspace, fred:looporder, sze:2020:epo}) the $S$ and $N$ ranks in the loop order and the rank order of $\operationinputmask$, calling for a new format for the $N$ and $S$ ranks (Figure~\ref{fig:opt-space:format}c).
After swizzling, we use an uncompressed format for the $N$ rank, which requires payloads to encode the number of operations associated with each operation type.
Additionally, because fibers of the $N$ rank are uncompressed (and therefore have a constant occupancy), the payloads of rank $I$ become unnecessary and are eliminated.
We continue to use a compressed format for the $S$ rank with coordinates only.
Payloads of the $S$ rank describe the occupancy of fibers in the $O$ rank.
However, this occupancy is already described by the $n$ coordinate, so the $S$ rank's payloads are redundant.

\subsection{Binding Optimization -- Loop Unrolling}
\label{sec:unrolling}

Loop unrolling~\cite{dragon-book} is a binding-level optimization that replicates the loop body based on 
knowledge of the loop bounds. 
Loop unrolling reduces loop overhead (e.g., the number of dynamic instructions) and exposes additional instruction scheduling opportunities across iterations---but at the cost of increased static code size. 

In \ourscheme, we design a sequence of progressively more optimized kernels.
That is, each kernel in the sequence implements all of its predecessors' optimizations plus new optimizations.
For example, OU includes the optimizations in RU, NU includes those in OU and RU, and so forth.

\para{R Rank Unrolling (RU)}
We start with a design that traverses the dataflow graph ($\operationinputmask$) using the format described in Figure~\ref{fig:opt-space:format}b.
Providing one extreme point in the spectrum of possible implementations, this kernel applies almost no loop unrolling, unrolling only the $R$ rank.
Fibers of the $R$ rank are always one-hot, so an $R$ loop would only add unnecessary overhead. 
Algorithm \ref{alg:RU} presents the pseudocode for RU.


\para{O Rank Unrolling (OU)}
Next, we completely unroll the $O$ rank.
This kernel uses the same loop order as RU.
Because the $O$ rank has no explicit metadata (see Figure~\ref{fig:opt-space:format}b), unrolling $O$ has no impact on the format of $\operationinputmask$.
However, unrolling does allow us to reduce redundant data movement (e.g., from $LI$ to $sel\_inputs$) and loop overheads.

\para{N Rank Unrolling (NU)}
To effectively unroll the $N$ rank, we first make a mapping-level change, swizzling the $S$ and $N$ ranks, producing an $[I, N, S, O, R]$ loop order.
This swizzling groups together outputs ($s$ coordinates) that are computed using the same operations ($n$ coordinates) in each layer.
We then fully unroll the $N$ rank, replacing the loop over the case statement with separate loops for each operation case body.
To maintain concordant traversal~\cite{teaal} of the $OIM$, we apply the corresponding swizzling and use the format described in Figure~\ref{fig:opt-space:format}c. Algorithm \ref{alg:NU} illustrates the NU implementation.
\begin{algorithm}
    \caption{NU kernel for the \ourscheme cascade}
    \label{alg:NU}
    \begin{algorithmic}[1]
        \small
        \State \textbf{for} $i = 0$ \textbf{to} $I-1$ do  \textbf{do} 
        \Comment{{\footnotesize Rank $I$}}
        \State \hspace{0.3cm} $n\_p = 0$
        \State \hspace{0.3cm} $\_, n\_p1 = OIM\_n$.next() \Comment{{\footnotesize Unrolled Rank $N$}}
        \State \hspace{0.3cm} $n\_p = n\_p + n\_p1$
        \State \hspace{0.3cm} \textbf{for} $s = 0$ to $n\_p1-1$ \textbf{do} \Comment{{\footnotesize Rank $S$}}
        \State \hspace{0.6cm}  $r\_c1 = OIM\_r$.next() \Comment{{\footnotesize Unrolled Rank $R$}}
        \State \hspace{0.6cm}  $r\_c2 = OIM\_r$.next() 
        \State \hspace{0.6cm} $LO[s] = LI[r\_c1] + LI[r\_c2]$ \Comment{{\footnotesize Evaluate add operation}}
        \State \hspace{0.3cm} $\_, n\_p2 = OIM\_n$.next()
        \State \hspace{0.3cm} 
        $n\_p = n\_p + n\_p2$
        \State \hspace{0.3cm} 
        $...$\Comment{{\footnotesize Evaluate other operations}}
        \State \hspace{0.3cm} \textbf{for} $s = 0$ \textbf{to} $n\_p-1$ \textbf{do} \Comment{{\footnotesize Rank $S$}}
        \State \hspace{0.6cm} 
        $s\_c, \_ = OIM\_s$.next()
        \State \hspace{0.6cm} 
        $LI[s\_c] = LO[s]$
        \Comment{{\footnotesize Write $LO$ back to $LI$}}
    \end{algorithmic}
\end{algorithm}

\para{Partial S Rank Unrolling (PSU)}
Next, we partially unroll the $S$ rank.
We unroll the $S$ loop of the last Einsum in 
Cascade~\ref{box:einsum} 24 times, while $S$ loops of other Einsums for the most common operators are unrolled 8 times.
(24 and 8 were chosen because they work well in practice.)
Because this is only partial unrolling, it has no impact on the format of $\operationinputmask$.

\para{I Rank Unrolling (IU)}
We then completely unroll the $I$ rank of all Cascade~\ref{box:einsum} Einsums.
This unrolling enables the elimination of zero-iteration $S$ loops, which arise when a given operation (i.e., an $n$ coordinate) is unused in a layer.

\para{S Rank Unrolling (SU)}
We now completely unroll the $S$ rank of all Cascade~\ref{box:einsum} Einsums, fully encoding $\operationinputmask$ in the binary and eliminating all associated metadata and loop overheads.

\para{Tensor Inlining (TI)}
Finally, we replace the array based representations of $\layerinput$ and $\layeroutput$ with individual 
C++ variables.
This transformation is analogous to function inlining.
We refer to it as \emph{tensor inlining}.
This kernel gives the  C++ compiler maximum flexibility to bind values to registers, reorder instructions, or eliminate them entirely when unnecessary.
\section{\ourscheme Proof-of-Concept Simulator}
\label{sec:compiler}

In this section, we present our proof-of-concept compiler and simulator, which extracts the tensors defined in Section \ref{sec:cascade} from an RTL design and evaluates them using the kernels described in Section \ref{sec:unrolling}. 
We further demonstrate how \ourscheme integrates with existing RTL design flows, including support for common features such as waveform generation.


\subsection{\ourscheme Design Overview}

\begin{figure}[t]
    \centering
    \includegraphics[width=0.9\linewidth]{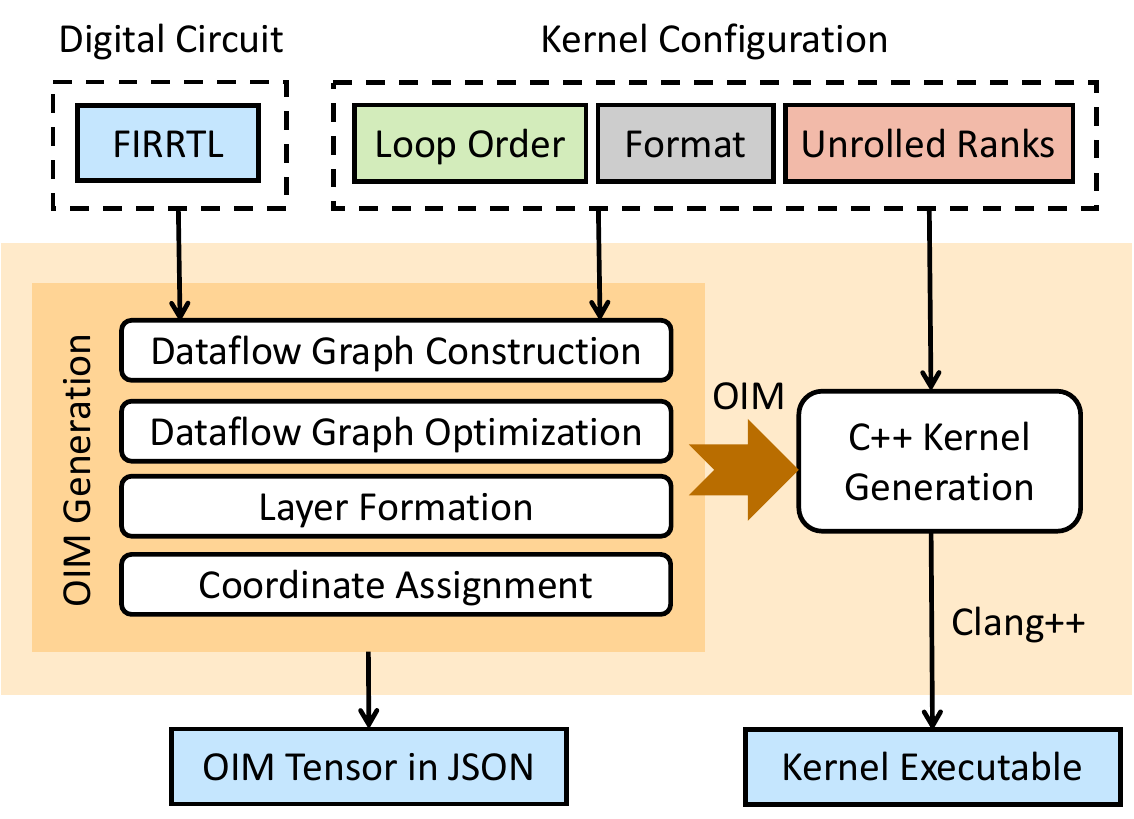} 
    \caption{\textbf{
    \ourscheme simulation overview.
    The compiler takes two inputs: (1) a digital circuit expressed in FIRRTL and (2) a kernel configuration.
    Through $OIM$ generation and C++ kernel generation, the compiler produces a kernel executable and the corresponding $OIM$ tensors stored in JSON files, which together achieve RTL simulation.
}}
    \label{fig:compiler}
\end{figure}

Figure~\ref{fig:compiler} provides a high-level overview of the \ourscheme simulation flow.
The compiler takes two inputs: (1) a digital circuit expressed in FIRRTL~\cite{firrtl}, and (2) a kernel configuration specifying the loop order, tensor format, and degree of unrolling (Section~\ref{sec:unrolling}). 
The end-to-end compilation workflow proceeds as follows.
First, the compiler extracts connectivity information from the FIRRTL input and constructs a dataflow graph of the design.
Next, the compiler applies a series of optimizations to the dataflow graph.
Some of these optimizations were used in prior RTL simulation work, including 
\emph{operator fusion}~\cite{essent-repo} (e.g., mux-chain extraction) and \emph{copy propagation}~\cite{essent,essent-repo}. 
More detail about these optimizations is available in Section \ref{sec:related-work} and 
Appendix~\ref{app:opt-explain}. 
We apply additional classical optimizations, e.g., constant propagation, as a means to optimize the $OIM$, as appropriate.

After optimization, the compiler slices the dataflow graph into layers via topological sort.
It then assigns (coordinate, payload) pairs for the $I$, $S$, $N$, $O$, and $R$ ranks and constructs the corresponding $OIM$ tensor according to the selected format (e.g., Figure~\ref{fig:opt-space:format}).
The $OIM$ tensor is stored in JSON files and loaded at runtime.
$OIM$'s $N$ rank supports 
all FIRRTL primitive operations~\cite{firrtl} and the
custom mux-chain operation.
The identity insertion and elision steps in Section~\ref{sec:cascade} are introduced for conceptual clarity in forming the cascade.
In the actual implementation, the compiler assigns the $s$ coordinates so that all identity operations can be elided.

Finally, the compiler generates a C++ simulation kernel.
Depending on the kernel configuration, the generated code corresponds to one of the seven kernels described in Section~\ref{sec:unrolling}.
The resulting C++ kernel is then compiled using \texttt{clang} to produce the executable used for simulation.

\subsection{System Integration}\label{sub:system-integration}
Next, we describe how \ourscheme 
supports common practical simulation requirements.

\para{HDL Support} \ourscheme takes FIRRTL~\cite{firrtl} as its input. FIRRTL is a widely used intermediate representation produced by modern hardware description languages (HDLs), like Chisel~\cite{chisel}, Spatial~\cite{Spatial} and PyRTL~\cite{pyrtl}.
Verilog designs can be translated to FIRRTL using tools such as Yosys~\cite{yosys}.

\para{Multiple Clock Domains}
\ourscheme targets circuits with a single clock domain.
Multi-clock designs can be supported by partitioning the circuit according to clock domain and adding a synchronization step at the end of each cycle.

\para{Host--DUT Communication}
RTL simulation typically requires communication between the host and the design under test (DUT) to load and run programs on the DUT.
As an example, to support the Debug Module Interface (DMI)~\cite{chipyard-dmi}, \ourscheme connects the frontend server (FESVR) and the DUT by reading and updating Debug Transfer Module (DTM) signals in the $LI$ at the end of each simulation cycle.

\para{Waveform Generation}
Waveform generation requires (1) exposing both internal and I/O signals and (2) recording signal values when they change.
To support this functionality, optimizations that eliminate signals are disabled.
Moreover, we assign unique $s$ coordinates to each signal to preserve its value across consecutive cycles.
By comparing each signal’s current value with its value in the previous cycle, \ourscheme detects signal transitions and can be used to generate waveforms.


\para{Cross-Module Referencing}
Cross-module referencing (XMR) allows an HDL module to reference signals defined in other modules, including non-I/O signals.
Both Chisel and Verilog support XMR. 
In fact, Chisel implements XMR via FIRRTL transformations that promote referenced signals to module I/O~\cite{circt-xmr}, producing a lowered FIRRTL representation that is directly accepted by \ourscheme.

\section{Evaluation}
\label{sec:eval}

In this section, we evaluate \ourscheme and compare it to prior state-of-the-art RTL simulators.
We emphasize that \ourscheme is a \emph{proof-of-concept} implementation, demonstrating the viability of tensor algebra as a foundation for RTL simulation rather than a fully optimized simulator.
Accordingly, our evaluation focuses on kernel design trade-offs, scalability, and showing that \ourscheme can achieve performance competitive with Verilator.

\subsection{Evaluation Set-Up}
\label{sec:eval:set-up}
We evaluate \ourscheme across a range of RTL designs, host machines, and kernel configurations.
Table~\ref{table:machine} summarizes the machines we use, spanning multiple ISAs (Arm and x86), vendors (Intel, AMD, and AWS), and platforms (desktop and server). 
We evaluate four RTL designs from Chipyard~\cite{chipyard}: RocketChip~\cite{rocketchip}, BOOM~\cite{zhaosonicboom}, Gemmini~\cite{gemmini-dac}, and SHA3~\cite{chipyard-sha3}.
RocketChip and BOOM run the \texttt{dhrystone} benchmark, Gemmini runs \texttt{matrix\_add-baremetal}, and SHA3 runs \texttt{sha3-rocc}.
Required simulation cycles for each design are reported in Table~\ref{table:simcycle}.
We compare against Verilator~5.016 and ESSENT, compiling with \texttt{clang -O3} unless otherwise noted.
All results are reported as the average of three runs.

\begin{table}[h]
\caption{\textbf{Summary of machines. All run Ubuntu and use clang++ 14.0.0 (Intel) or 18.1.3 (AMD and AWS).}}
\label{table:machine}
\centering
\footnotesize  
\begin{tabular}{>{}l l >{}l l}
\hline
\multicolumn{2}{c}{\textbf{Intel Core i9-13900K}} & \multicolumn{2}{c}{\textbf{Intel Xeon Gold 5512U}} \\
\hline
L1 I/D-Cache & 32/48 KB & L1 I/D-Cache & 32/48 KB  \\
L2 Cache & 2 MB & L2 Cache & 2 MB \\
LLC      & 36 MB & LLC & 52.5 MB \\
\hline
\multicolumn{2}{c}{\textbf{AMD Ryzen 7 4800HS}} & \multicolumn{2}{c}{\textbf{AWS Graviton 4}} \\
\hline
L1 I/D-Cache & 32/32 KB  & L1-I/D Cache & 64/64 KB \\
L2 Cache & 512 KB & L2 Cache &  2 MB \\
LLC      & 8 MB & LLC & 36 MB \\
\hline
\end{tabular}
\end{table}

\begin{table}[h]
\centering
\footnotesize
\caption{\textbf{Simulation cycles reported in thousands (K) and rounded to two significant figures for different RTL designs.
Gemmini’s simulation cycles vary significantly across different matrix configurations (8×8, 16×16, and 32×32).}}
\label{table:simcycle}
\begin{tabular}{lrrrr}
\toprule
RTL Designs & RocketChip & BOOM & Gemmini-8/16/32 & SHA3 \\
\midrule
Sim Cycles (K) & 540 & 750 & 160/350/1100 & 1200 \\
\bottomrule
\end{tabular}
\end{table}


\subsection{Ablation Study 1: Kernel Configurations}\label{subsec:study1}
We compile and simulate the 8-core RocketChip on all four machines using different kernels to study the trade-offs between kernel configurations for a fixed design.

\begin{figure}[h]
    \centering
\includegraphics[width=0.9\linewidth]{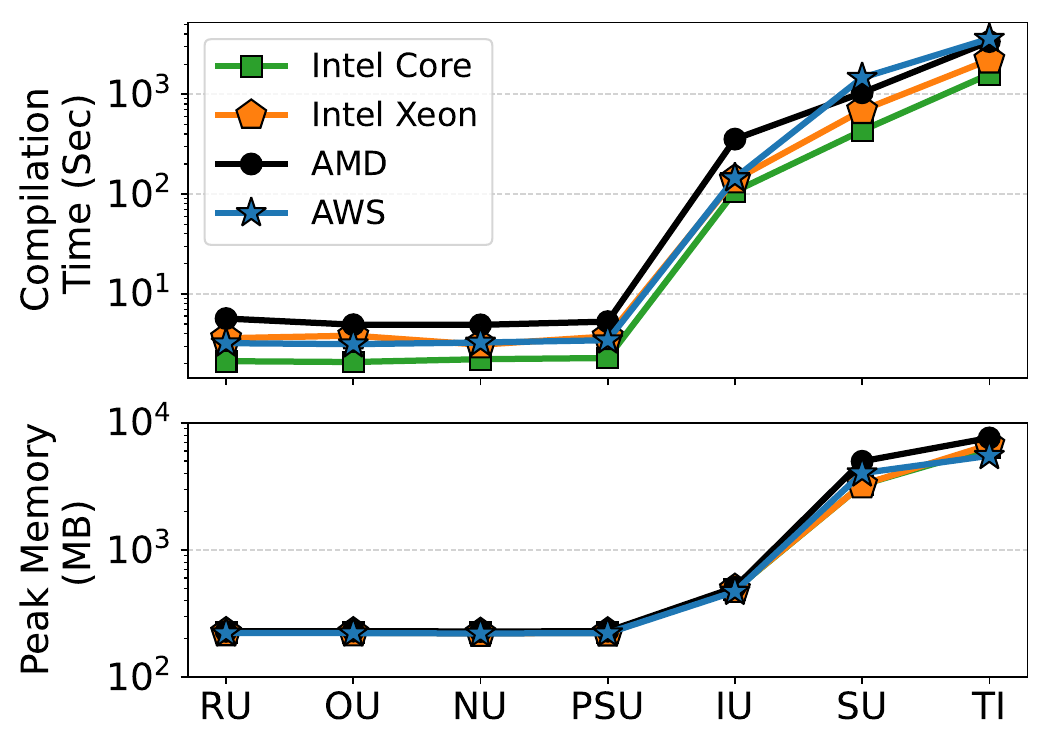} 
    \caption{
    \textbf{Compilation time (top) and peak memory (bottom) for \ourscheme when compiling 8-core RocketChip.
    }}
    \label{fig:compile}
\end{figure}

\para{Compilation Comparison}
Figure~\ref{fig:compile} shows compilation time and peak memory usage for different \ourscheme kernels. 
As we move from the mostly rolled kernel (RU, common in tensor algebra) to the fully unrolled kernel (TI, typical in RTL simulation), both compilation time and memory usage increase. 
This trend stems from the growing binary size, as also shown in Table~\ref{table:binsize}.
Larger binaries demand more compiler resources for aggressive optimizations, leading to longer compile times and higher memory consumption.

\begin{table}[h]
\centering
\caption{\textbf{Binary size of \ourscheme kernels on Intel Xeon.
}}
\label{table:binsize}
\footnotesize
\begin{tabular}{lccccccc}
\toprule
Kernels & RU & OU & NU & PSU & IU & SU & TI \\
\midrule
Size (MB)  
 & 0.35 & 0.35 & 0.34 & 0.35 & 0.91 & 6.0 & 5.3 \\
\bottomrule
\end{tabular}
\end{table}

\para{Simulation Analysis} 
Next, we examine simulation performance across kernels.
Unrolling kernels introduces competing performance effects.
As shown in Table~\ref{table:instr},  unrolling reduces the dynamic instruction count by eliminating loop overhead. 
However, excessive unrolling increases the static code size, which leads to higher I-cache miss rates and lower IPC. 
A similar trade-off arises between D-cache and I-cache pressure.
In the mostly rolled implementation, the entire 
$OIM$ is represented as 
arrays and stored in the D-cache. 
As the kernel is progressively unrolled, we gradually embed the $OIM$ into the binary, shifting pressure from the D-cache to the I-cache.
These trends are illustrated in Table~\ref{table:cache}. 
Additionally, while L1 D-cache load counts drop drastically from RU to SU, the miss counts remain relatively stable.
The $OIM$ accesses are mostly sequential, allowing them to be efficiently handled by the stride prefetcher, even when $OIM$ is mainly stored as data arrays in rolled kernels.
In contrast, the $LI$ tensor, although smaller than the $OIM$, is accessed irregularly, making it the primary source of D-cache misses.

\begin{table}[t]
\centering
\footnotesize
\caption{\textbf{Dynamic instructions reported in trillions (T) and IPC for 8-core RocketChip simulated on Intel Xeon.}}
\label{table:instr}
\begin{tabular}{lrrrrrrr}
\toprule
Kernels & RU & OU & NU & PSU & IU & SU & TI \\
\midrule
Dyn. Inst. (T) & 26.9 & 2.79 & 1.33 & 1.24 & 1.31 & 0.539 & 0.476 \\
IPC            & 4.36 & 3.91 & 2.93 & 2.73 & 2.47 & 0.59  & 0.66 \\
\bottomrule
\end{tabular}
\end{table}

\begin{table}[t]
\centering
\footnotesize
\caption{\textbf{Instruction and data cache profiling results reported in billions (B) for 8-core RocketChip simulated on Intel Xeon.
}
}
\label{table:cache}
\begin{tabular}{lrrrrrrr}
\toprule
Kernels & RU & OU & NU & PSU & IU & SU & TI \\
\midrule
L1I Miss (B) & 0.28 & 0.04 & 0.03 & 0.03 & 5.08 & 50.81 & 39.17 \\
L1D Load (B) & 8190 & 910.2 & 619.6 & 620.4 & 649.2 & 241.2 & 196.4 \\
L1D Miss (B) & 71.43 & 69.78 & 73.36 & 69.01 & 68.96 & 76.62 & 39.17 \\
\bottomrule
\end{tabular}
\end{table}

\begin{figure}[h]
    \centering
\includegraphics[width=0.9\linewidth]{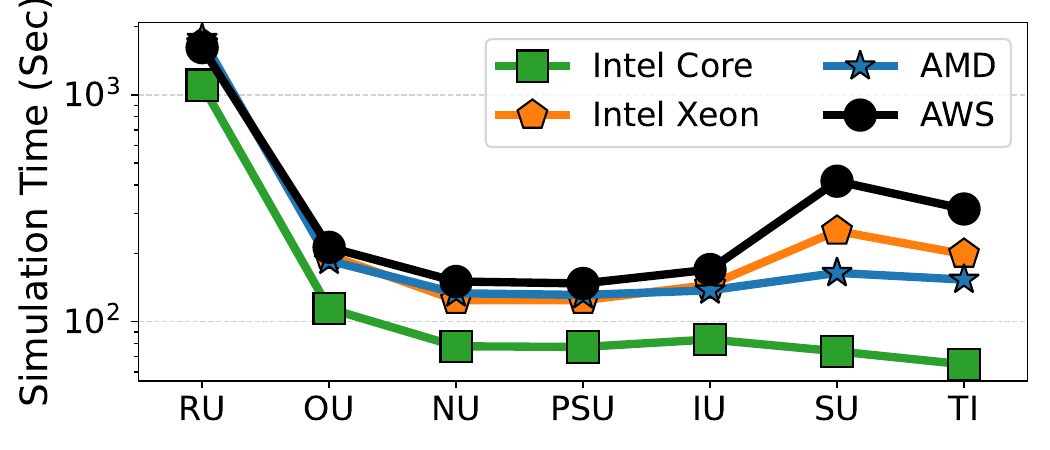} 
    \caption{
    \textbf{Simulation time of \ourscheme kernels when simulating 8-core RocketChip.
    }}
    \label{fig:r8_perf}
\end{figure}

These observations raise the question of whether there exists a ``sweet spot” that (1) balances reduced dynamic instructions against IPC loss, and (2) balances I-cache and D-cache pressure. 
Figure~\ref{fig:r8_perf} answers this question affirmatively, showing simulation performance across all seven kernel configurations.
We make two observations from Figure~\ref{fig:r8_perf}.

\para{Existence of a sweet spot}
We first notice that a sweet spot exists in the middle of the unrolling spectrum. 
Across the Intel Xeon, AMD, and AWS machines, the PSU kernel achieves the highest performance.
Top-down analysis reveals why partially rolled kernels can outperform fully unrolled ones.
For instance, on the Intel Xeon, the frontend-bound fraction for rolled kernels such as NU and PSU is around 5\%, but it jumps to roughly 80\% for the fully unrolled SU kernel.
A similar trend occurs on both the AMD and ARM platforms.

\para{Best kernel varies by machines}
Second, the most performant kernel for the same design can differ across machines.
Unlike the other three machines, TI performs best on the Intel Core. 
Top-down analysis shows that frontend-bound behavior drives this difference: highly unrolled kernels (SU and TI) are about 80\% frontend-bound on the Intel Xeon but only 15–25\% on the Intel Core.
As a result, unrolled kernels are more favorable on the Intel Core, since they execute fewer dynamic instructions and the frontend can sustain the increased pressure.
Yet, as design size increases, rolled kernels will outperform unrolled ones even on the Intel Core.

The different frontend-bound behavior between the Intel Xeon and Core are primarily due to fetch latency.
On Intel Xeon, fetch latency accounts for over 90\% of frontend stalls, which correlates with the fact that its last-level cache latency is roughly twice that of the Intel Core~\cite{xeon6-submem}.

\subsection{Ablation Study 2: Scalability Across Design Size}
\label{subsec:study2}


Next, we evaluate the scalability of \ourscheme kernels by increasing the size of the design while keeping other factors constant.
All experiments in this section are performed on the Intel Xeon platform.

\begin{figure}[h]
    \centering
\includegraphics[width=0.9\linewidth]{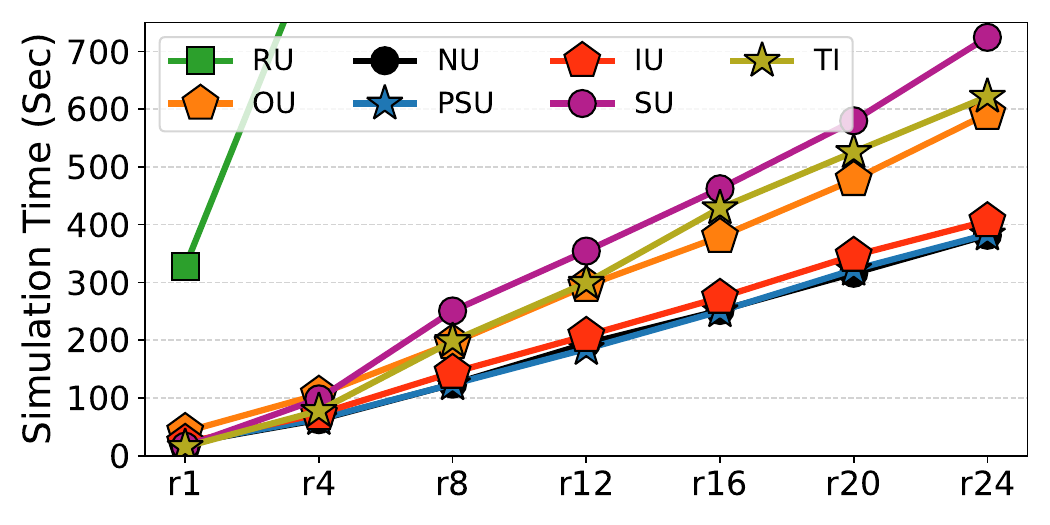} 
    \caption{
    \textbf{Simulation time of \ourscheme kernels when simulating 1 to 24-core RocketChips (r1-r24) on Intel Xeon.
    }}
    \label{fig:kernels}
\end{figure}

\para{Kernel performance with increasing design size}
We simulate 1 to 24-core RocketChips 
(FIRRTL sizes 18–150+ MB) across seven kernel configurations.
Figure~\ref{fig:kernels} presents the results. 
The performance of the RU kernel is significantly worse than the other kernels, so only its first data point is shown for clarity.
The results show that the optimal kernel can change with design size, even on the same machine. 
For example, the TI kernel performs best on the 1-core RocketChip, but as the design scales, it becomes increasingly limited by frontend stalls. 
When scaling from 1 to 24 cores, the frontend stall rate increases from 48.2\% to 81.4\%.
In contrast, partially rolled kernels such as NU and PSU scale better, exhibiting near-linear performance and outperforming TI from the 4-core design onward.
PSU’s frontend-bound rate, for instance, increases only slightly, from 6.2\% on the 1-core design to 6.5\% on the 24-core design.
Moreover, as shown in Section~\ref{subsec:study1}, rolled kernels produce small binaries and incur short compilation time (on the order of a few seconds), making them scalable in both compilation and simulation.

\para{Comparing Scalability with Prior Work}
We select the partially unrolled PSU kernel, which demonstrated good scalability, to compare against Verilator and ESSENT in compilation overhead and simulation performance.
Compiling ESSENT with \texttt{clang -O3} requires substantial memory and exceeds the DRAM capacity of our Intel Xeon Gold 5512U machine.
We therefore perform ESSENT compilation on a separate Intel Xeon machine (Gold 6248) with 256~GB of memory, and execute the binary on the Intel Xeon Gold 5512U.
Table~\ref{table:compile_overhead} reports the compilation time and memory usage for Verilator, ESSENT, and \ourscheme's PSU kernel.
PSU exhibits a significantly lower and nearly constant compilation cost as design size increases.
In contrast, compilation times of ESSENT and Verilator
grow with design size, and ESSENT incurs much higher overhead than Verilator.



\begin{table}[h]
\centering
\footnotesize
\caption{\textbf{Compilation overhead for Verilator, ESSENT, and \ourscheme's PSU kernel when compiling 1 to 24-core RocketChips (r1-r24) on Intel Xeon Gold 6248.
}}
\label{table:compile_overhead}


\begin{subtable}{\columnwidth}
\centering
\caption{Compilation time (seconds)}
\label{table:compiletime}
\begin{tabular}{lrrrrrrr}
\toprule
Design & r1 & r4 & r8 & r12 & r16 & r20 & r24 \\
\midrule
Verilator & 92.0 & 164 & 277 & 402 & 522 & 627 & 724 \\
ESSENT    & 121  & 582 & 1700 & 3980 & 7070 & 11000 & 13700 \\
PSU & 4.26 & 4.26 & 4.26 & 4.26 & 4.26 & 4.26 & 4.26 \\
\bottomrule
\end{tabular}
\end{subtable}


\begin{subtable}{\columnwidth}
\centering
\caption{Compilation peak memory usage (GB)}
\label{table:compilememory}
\begin{tabular}{lrrrrrrr}
\toprule
Design & r1 & r4 & r8 & r12 & r16 & r20 & r24 \\
\midrule
Verilator & 0.227 & 0.257 & 0.350 & 0.368 & 0.267 & 0.276 & 0.282 \\
ESSENT    & 2.80  & 11.9  & 34.2  & 67.2  & 114   & 168   & 234 \\
PSU & 0.203 & 0.203 & 0.203 & 0.203 & 0.203 & 0.203 & 0.203 \\
\bottomrule
\end{tabular}
\end{subtable}
\end{table}

\begin{figure}[h]
    \centering
\includegraphics[width=0.9\linewidth]{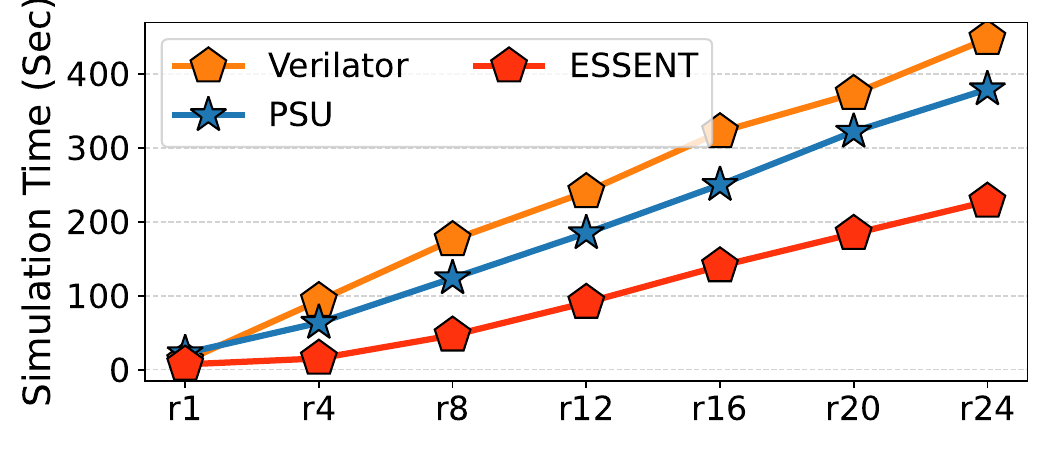} 
    \caption{
    \textbf{Simulation time of Verilator, ESSENT, and \ourscheme's PSU kernel when simulating 1 to 24-cores RocketChips (r1-r24) on Xeon.
    }}
    \label{fig:PSU}
\end{figure}

Figure~\ref{fig:PSU} shows simulation performance for the three simulators compiled with \texttt{clang -O3}.
All scale with design size, but Verilator exhibits the longest simulation times, the PSU kernel is moderately faster, and ESSENT achieves the best performance.
ESSENT’s advantage comes from two main factors. 
First, unlike Verilator, ESSENT generates fully straight-line code, reducing branch misprediction overhead: for 4-core RocketChip, Verilator’s branch miss rate is 22\%, whereas ESSENT’s is only 0.1\% (0.12\% for the PSU kernel).
Second, straight-line code allows \texttt{clang} to apply more aggressive optimizations, so ESSENT executes far fewer instructions than Verilator or the PSU kernel.
The downside is extremely high compilation cost, which grows rapidly with design size.
In the next study, we examine the impact of compiler optimizations on these simulators.

\subsection{Ablation Study 3: C++ Compiler Effects}
So far, all simulators are compiled with \texttt{clang -O3}.
Understanding the impact of compiler optimizations is important, since aggressive optimizations can incur high compilation overhead.
To assess baseline performance and dependence on compiler optimizations, we evaluate simulations compiled with \texttt{clang -O0}.

\begin{figure}[h]
    \centering
\includegraphics[width=0.9\linewidth]{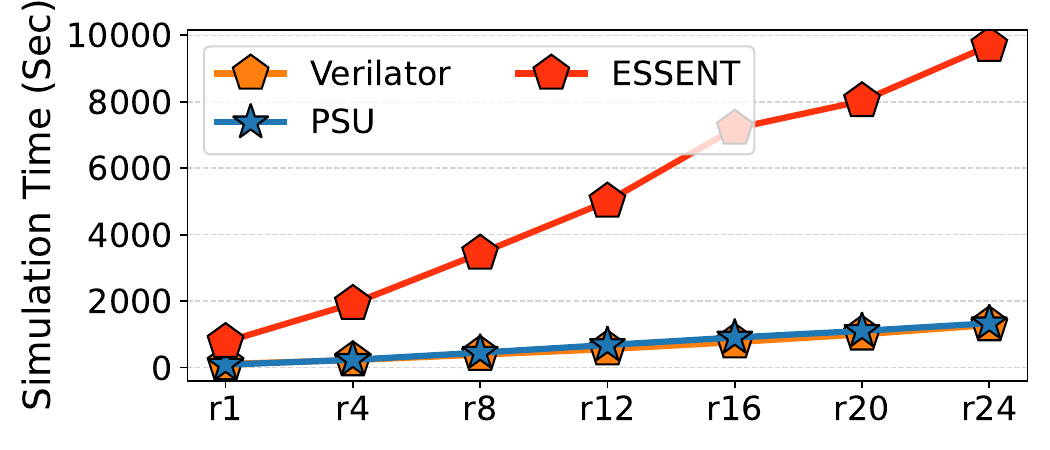} 
    \caption{
    \textbf{Simulation time of Verilator, ESSENT, and \ourscheme's PSU kernel when running 1- to 24-core RocketChips (r1-r24) on Intel Xeon, compiled with \texttt{-O0}.
    }}
    \label{fig:O0}
\end{figure}


Figure \ref{fig:O0} shows the simulation performance of Verilator, ESSENT, and the PSU kernel on the Intel Xeon machine when compiled with \texttt{clang -O0}.
Our kernel and Verilator exhibit comparable performance, whereas ESSENT suffers a severe degradation.
The drastic slowdown of ESSENT is primarily due to a much larger dynamic instruction count in the unoptimized program.
Compared with the \texttt{-O3} version, the dynamic instruction count increases on average by 3.8$\times$ for PSU, 4.42$\times$ for Verilator, and 103.3$\times$ for ESSENT.
ESSENT relies on straight-line code aggressively optimized by the compiler; when optimizations are disabled, this advantage largely disappears, making ESSENT highly compiler-dependent.

\subsection{Main Evaluation}
\begin{figure*}[th]
    \centering
    \includegraphics[width=\textwidth]{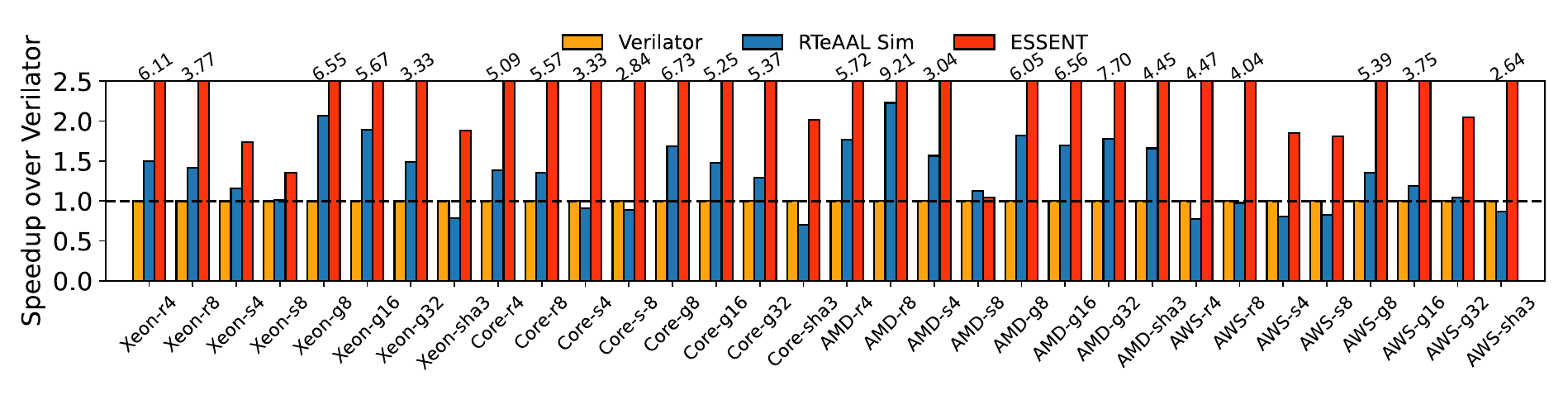}
    \caption{\textbf{Simulation speedup of \ourscheme and ESSENT relative to Verilator across different RTL designs and host machines. X-axis prefixes \textit{r-}, \textit{s-}, and \textit{g-} denote different sizes of RocketChip, SmallBOOM, and Gemmini designs, respectively.
}}
    \label{fig:speedup}
\end{figure*}

Lastly, we evaluate all seven \ourscheme kernels across all RTL designs (Section~\ref{sec:eval:set-up}) on four machines, compiling all with \texttt{clang -O3}.
For each design and machine, we report the simulation time of the best-performing \ourscheme kernel and compare it against Verilator and ESSENT in Figure~\ref{fig:speedup}.
Across all designs, \ourscheme achieves performance comparable to Verilator, with speedups observed on every machine.
From Figure~\ref{fig:speedup}, we make three observations.

\para{The LLC effects on performance}
Last-level cache (LLC) capacity can be a dominant factor in simulation performance, especially as binary size grows.
When simulating the 8-core SmallBOOM design on the AMD machine, \ourscheme achieves the best performance---the only simulation setting where it outperforms both Verilator and ESSENT.
The combination of a small LLC (8MB) and large simulator binaries (11MB for ESSENT, 19~MB for Verilator) causes ESSENT’s fully unrolled straight-line code to exceed LLC capacity, leading to frequent LLC misses.
In contrast, \ourscheme's rolled kernels generate compact binaries that fit within the LLC, delivering superior performance under tight cache constraints.

\begin{figure}[h]
    \centering
\includegraphics[width=0.9\linewidth]{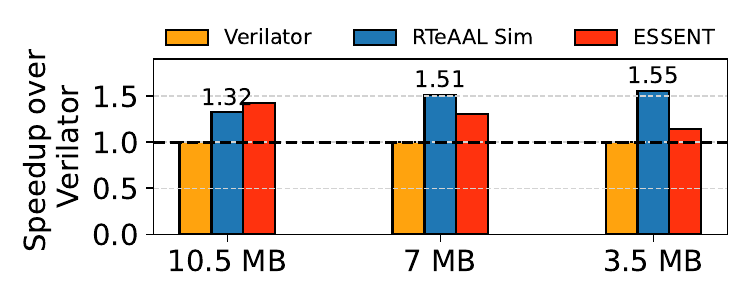} 
    \caption{\textbf{Speedup of \ourscheme (PSU kernel) and ESSENT relative to Verilator  for 8-core SmallBOOM as LLC capacity decreases (X-axis) on the Intel Xeon.
    }}
    \label{fig:LLC}
\end{figure}

To validate this, we use Intel Cache Allocation Technology (CAT) to restrict the LLC on the Intel Xeon.
Figure~\ref{fig:LLC} reports the speedup of \ourscheme for the 8-core SmallBOOM design with 10.5MB, 7MB, and 3.5MB LLCs.
As LLC capacity decreases, ESSENT’s performance drops sharply, consistent with prior observations~\cite{dedup}.
In contrast, \ourscheme's PSU kernel maintains stable performance, again demonstrating good scalability under constrained cache resources.

\para{Performance variation on different machines}
Differences in the hardware and software stack can affect the relative performance of RTL simulators.
Among four host machines,
\ourscheme performs worst relative to Verilator on the AWS Graviton 4.
Empirically, we observe that Verilator’s branch misprediction rate is much lower on AWS than on x86 machines.
For example, when simulating the 4-core RocketChip, Verilator’s branch misprediction rate drops from 22\% on Intel Xeon to 0.22\% on AWS.
In contrast, \ourscheme (and ESSENT) already maintain low misprediction rates on x86, so the reduced misprediction on AWS adds little benefit.
We do not have sufficient evidence to attribute this behavior to a single root cause.
It may stem from microarchitectural differences (e.g., branch predictor), compiler effects (e.g., code layout produced by the ARM and x86 backends), or their interaction.

\para{Performance variation across RTL designs}
\ourscheme consistently outperforms Verilator on all RTL designs except SHA3. 
This is because the current prototype performs best for designs that are cache-bound, where the rolled kernels benefit from reduced frontend pressure. 
However, because SHA3 is a relatively small design, its most performant kernel is TI. 
TI is a straight-line kernel similar to prior simulators, where C++ variables are used (instead of arrays) to store signal values. 
Since we do not apply optimizations as aggressive as Verilator, we cannot achieve competitive performance in this case. 

\section{Related Work}
\label{sec:related-work}

\newtcolorbox[auto counter]{relatedworkbox}[2][]{%
   colback=white,
  colframe=black!15,
  colbacktitle=blue!10,
  coltitle=black,
  sharp corners,
  boxrule=0.5pt,
  fonttitle=\bfseries,
  arc=2pt,
  center title,
  title=Box \thetcbcounter: #2,
  label={#1}
}

\begin{figure}
\centering
\begin{relatedworkbox}[box:RTL-TeAAL]{RTL Simulation Optimizations in TeAAL}
\footnotesize
\paragraph{Cascade}
\begin{itemize}[leftmargin=1em, noitemsep, topsep=0pt]
    \item Skipping partitions w/o activity~\cite{essent}
    \item Synchronization of partitions~\cite{repcut, parendi, manticore}
    \item Differential exchange (send only changed values)~\cite{parendi}
    \item Instance reuse~\cite{dedup}
    \item \textbf{Operator fusion (mux/or/xor chains)}~\cite{essent-repo}
    \item ML for RTL simulation ~\cite{MLmodel}
\end{itemize}

\paragraph{Mapping}
\begin{itemize}[leftmargin=1em, noitemsep, topsep=0pt]
    \item Hypergraph partitioning~\cite{repcut, gem}
    \item Traversal exploiting temporal locality~\cite{dedup}
    \item Parallelize across partitions~\cite{repcut, parendi, gem, Metro-Mpi}
\end{itemize}

\paragraph{Format}
\begin{itemize}[leftmargin=1em, noitemsep, topsep=0pt]
    \item Optimize memory layout~\cite{repcut}
    \item \textbf{Compress dataflow graph representation}~\cite{gem}
\end{itemize}

\paragraph{Binding}
\begin{itemize}[leftmargin=1em, noitemsep, topsep=0pt]
    \item Branch hints~\cite{essent}
    \item \textbf{Function inlining}~\cite{parendi}
    \item Software pipelining ~\cite{ToraRTL}
    \item Multi-platform task scheduling ~\cite{cuda}
\end{itemize}

\paragraph{Data}
\begin{itemize}[leftmargin=1em, noitemsep, topsep=0pt]
    \item Replicate/partition for parallelism~\cite{repcut, parendi, gem}
    \item \textbf{Copy propagation}~\cite{essent, essent-repo}
\end{itemize}
\end{relatedworkbox}
\end{figure}


\para{Optimizations by prior RTL simulators} Prior work has optimized RTL simulation for CPUs~\cite{essent, essent-mag, essent-repo, repcut, dedup, Metro-Mpi}, GPUs~\cite{gem, cuda}, and accelerators~\cite{parendi, rtl-sim-hw-sw-codesign, manticore}.
We view these efforts as complementary to \ourscheme, and show that their optimizations can be incorporated into \ourscheme by mapping them onto the TeAAL hierarchy.
In Box~\ref{box:RTL-TeAAL}, we classify 17 techniques from 11 prior works and map them onto the TeAAL hierarchy, with details in 
Appendix~\ref{app:opt-explain}.

As an example, RepCut~\cite{repcut} proposes a hypergraph partitioning algorithm that enables multithreaded simulation with low replication overhead.
This algorithm can be formulated as a modification to the cascade (see 
Appendix~\ref{app:repcut-einsum}). 

Some techniques, such as copy propagation, modify the dataflow graph (or, in our case, the values in the $OIM$ tensor).
To capture these techniques in TeAAL, we extend the TeAAL hierarchy with a new \textbf{\emph{data level}}, which represents transformations applied directly to the underlying tensors. 
Optimizations shown in \textbf{bold} are implemented in our current prototype. 
Appendix~\ref{app:opt-explain} 
describes how these optimizations are selected and realized in \ourscheme.

\para{RTL Simulation on Advanced Hardware Platforms} 
In this work, we focus on CPU-based simulation, as CPUs remain the dominant platform for state-of-the-art compilation-based RTL simulators.
Yet, the tensor-algebraic formulation naturally generalizes to other platforms, including GPUs and custom ASICs.
Within the TeAAL hierarchy, the target architecture constrains the design space expressible at each level.
As a result, different architectures may call for a different combination of optimizations to achieve high performance.

For example, a GPU backend expands the mapping space by enabling fine-grained parallelism.
GEM~\cite{gem} exploits this by partitioning the dataflow graph (mapping level) into fine-grained sectors scheduled across threads.
To mitigate thread divergence and irregular memory access, GEM rewrites circuits into an And-Inverter Graph (AIG) consisting only of AND and INV gates (data level); this uniform data representation enables the mapping of logic onto its 
virtual Very Long Instruction Word (VLIW) architecture for efficient parallel execution on GPUs.
These optimizations are particularly well suited to GPUs and can be viewed as architecture-specific.
Similar extensions are applicable to other accelerator architectures~\cite{rtl-sim-hw-sw-codesign, parendi, manticore}. 
We view these architecture-aware adaptations as promising directions for future work.

\para{Tensor Algebra for Applications from Other Domains} 
A growing body of work in the tensor algebra community has explored using tensor algebraic formulations to represent workloads from adjacent domains ~\cite{teaal, fusemax, edge, continuous-tensor-abstraction, graphblas, graphmat, EinHops, SecureLoop}. 
For example, prior work has expressed graph algorithms using tensor algebra abstractions, such as GraphBLAS~\cite{graphblas} and GraphMat~\cite{graphmat}. 
Other efforts express graph algorithms with recursive structure (e.g., Bellman–Ford) by supporting tensor recurrences~\cite{Recurrence}, similar to the iterative rank used in this work. 
Beyond EDGE, other work has also explored extending tensor algebra expressivity through user-defined operators~\cite{Olivia}.
Together, this literature highlights the generality and expressiveness of tensor algebra, and its potential to model a broad range of applications, beyond RTL simulation.

\section{Conclusion and Future Work}

This paper proposes \ourscheme, a reformulation of RTL simulation as a sparse tensor algebra problem.
By expressing RTL simulation using the extended Einsum notation and leveraging the TeAAL hierarchy for systematic optimization, \ourscheme addresses key limitations of state-of-the-art compilation-based simulators, most notably high compilation cost and frontend bottlenecks.
Our prototype simulator demonstrates that even with a few tensor algebra optimizations, \ourscheme can mitigate these bottlenecks and achieve simulation performance competitive with Verilator. 

Beyond further optimizing \ourscheme on CPUs, we see significant potential in co-designing \ourscheme with hardware accelerators for sparse tensor algebra.
Existing ASICs and proposed hardware accelerators for RTL simulation~\cite{palladium, rtl-sim-hw-sw-codesign} are specialized exclusively for RTL workloads, which limits their demand, fabrication volume, and broader adoption.
In contrast, there is substantial ongoing effort to design~\cite{sam, stellar, extensor} and fabricate~\cite{onyx} general-purpose sparse tensor algebra accelerators, driven by a wide range of applications~\cite{sze:2020:epo, mahmoud:2020:tde, albericio:2016:cin, graph_clustering, nagasaka2019performance, graphblas, triangle_counting_tensor_algebra, tallskinny0, tallskinny1, vande:2012:lss, wilhelm:2016:lsc, hutter:2014:cas}.
By extending such accelerators with EDGE operators, they could in principle accelerate \ourscheme as well, pointing to a path toward commoditized, widely deployable, and reusable hardware acceleration for RTL simulation.



\begin{acks}
We thank the anonymous reviewers for their insightful and constructive feedback. 
This research was funded by NSF grants CNS-1954521,
CNS-1942888,
CNS-2154183 and
CCF-8191902; as well as by a gift from Annapurna Labs/Amazon AWS.
We would like to thank Toluwanimi O. Odemuyiwa, Joel S. Emer and Michael Pellauer for many helpful discussion on the cascade construction for \ourscheme.
We would also like to thank Haoyuan Wang and Scott Beamer for their assistance with the ESSENT, RepCut, and Dedup infrastructure. 
Finally, we thank Tianrui Wei for his help with the simulator infrastructure and for feedback at early stages of this work.
\end{acks}

\bibliographystyle{ACM-Reference-Format}
\balance
\bibliography{refs}
\newpage
\appendix



\section{Explanation for the Extra Ranks in Einsum \ref{eq:step5b}}\label{app:ostar}
The additional ranks, most notably the $O*$ rank on the left-hand side of Einsum~\ref{eq:step5b}, are introduced by the populate-coordinate operator.
Unlike the other operators, which operate on individual points and values, the populate-coordinate operator acts on an entire fiber of the output at a time.
Based on the given reduce temporary and the current output fiber, this operator defines which points in the output fiber to update and which to delete.

For example, consider a tensor $A$ with one rank $R$.
If the goal is to produce an output tensor containing the two largest values of $A$, we can define a custom populate-coordinate operator, denoted 
$max2$.
The corresponding Einsum is:
\begin{align}\label{eq:populate}
\scalebox{1}{$
B_{r*} = A_{r} ::  \lll \mathbbm{1} (max2)
$}
\end{align}

\begin{figure}[h]
    \centering
    \includegraphics[width=0.9\linewidth]{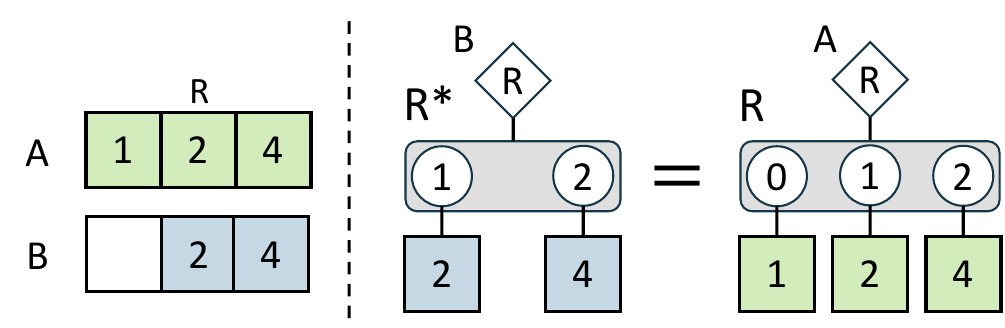}
    \caption{\textbf{Tensors involved in Einsum \ref{eq:populate} and their associated fibertrees.}}
    \label{fig:populate}
\end{figure}

Figure \ref{fig:populate} shows the tensors and corresponding fibertrees for this Einsum. 
As the figure shows, we "select" the largest two values from the R-fiber,
with the output tensor preserving the $R$ rank. 
We use the rank variable expression $r*$ to specify the populate-coordinate operator acts on the R-fiber.

\begin{figure}[hb]
    \centering
    \includegraphics[width=0.6\linewidth]{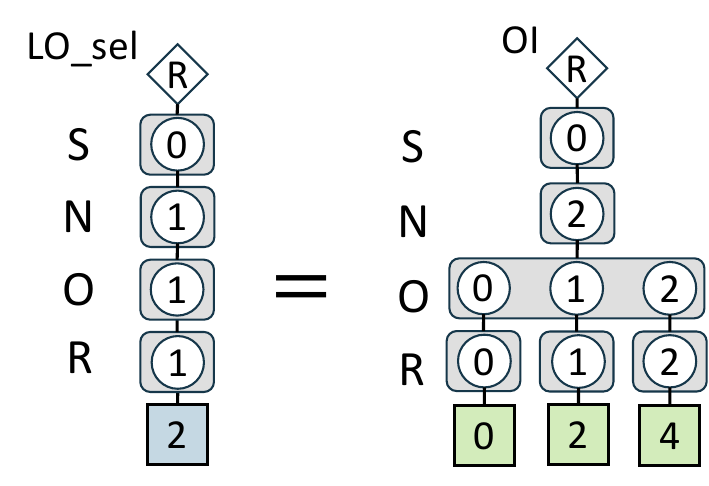}
    \caption{\textbf{Fibertree notations of tensors in Einsum \ref{eq:step5b}.}}
    \label{fig:populat_o}
\end{figure}

This is why we have $o*$ in $LO\_sel$.  We apply the populate coordinate operator on the O-fibers. Recall the Einsum \ref{eq:step5b}: 
\[
\scalebox{1}{$
\layeroutput\_sel_{n, o*, r, s} = OI_{n, o, r, s} ::  \bigwedge \mathbbm{1} (\leftarrow) \lll \mathbbm{1} (op\_s[n])
$}
\]
The operator $op\_s[n]$ effectively implements a multiplexer: based on the leaf payload value at the point with $O = 0$, it selects either the point with $O = 1$ or $O = 2$.
Consequently, the $O$ rank is preserved, as are all other ranks, since no reduction occurs.
Figure \ref{fig:populat_o} shows example fibertrees for this Einsum.



\section{Box~\ref{box:RTL-TeAAL} Optimization Explanations}\label{app:opt-explain}
In this section, we provide a detailed explanation of the optimizations summarized in Box~\ref{box:RTL-TeAAL} and their correspondence to the different levels of the TeAAL hierarchy. 
While changes at a higher level of the hierarchy can affect the lower levels due to their dependency, we assign each optimization to the level that most closely reflects its definition, typically the most abstract (highest) level it can influence.

\subsection{Implemented Optimizations}\label{subsec:impl-opt}
We first introduce the four optimizations we implemented in \ourscheme.
These were selected to cover different levels of the TeAAL hierarchy and to demonstrate both their applicability and how they manifest in the context of RTL simulation.
Furthermore, we prioritized techniques that synergize effectively with loop unrolling, a central optimization in our framework.
When unrolling is minimal, performance is primarily limited by the number of dynamic instructions, and all four optimizations contribute to reducing this count.
At maximum unrolling, performance becomes constrained by I-cache misses; in this regime, all optimizations except the \emph{compressed dataflow graph representation} help reduce binary size, thereby lowering I-cache pressure.

\para{Operator fusion ~\cite{essent-repo}}
Real world digital circuits contain repeated structures (e.g., mux chains) that can be fused and evaluated as a single larger operation to reduce the number of memory accesses required.
Changing the operations performed by the kernel changes \emph{what} compute is performed and, therefore, is a cascade-level optimization.
In our prototype, we apply this optimization during the dataflow graph optimizations phase in Figure~\ref{fig:compiler}.

\para{Compressed dataflow graph  ~\cite{gem}} 
Compressing tensors changes how they are laid out in memory to
reduce cache capacity requirements and memory traffic costs.
We therefore classify compression as a format-level optimization.
This optimization is described in detail in Section~\ref{sec:format}.

\para{Function inlining ~\cite{parendi}}
Inlining function calls eliminates call overhead and exposes additional opportunities for compiler optimization.
Inlining transforms the C++ program, so is a binding-level optimization.
Our prototype applies this optimization during C++ kernel generation, where all function calls are inlined to reduce the dynamic instruction count.

\para{Copy propagation ~\cite{essent, essent-repo}}
Eliminating redundant values from intermediate signals changes the dataflow graph, and is therefore classified as a data-level optimization.
We integrate this optimization in the dataflow graph optimizations phase in Figure~\ref{fig:compiler}.

\subsection{Cascade Level Optimizations}
\para{Skipping partitions without activity ~\cite{essent}}  
Prior work, such as ESSENT with \texttt{O3} optimization, implements activity-aware simulation.
The core idea is to partition the dataflow graph into smaller segments and track signal changes for each partition.
At each cycle, only partitions whose input signals have changed are simulated.
This approach fundamentally modifies the kernel algorithm, requiring the cascade to account for both signal recording and the conditional execution of each partition.



\para{Synchronization of partitions ~\cite{repcut, parendi, manticore}} 
In multithreaded RTL simulation, the synchronization step requires augmenting the cascade with additional Einsums.
For example, Appendix~\ref{app:repcut-einsum} demonstrates how a new Einsum is incorporated into the cascade to represent the RepCut~\cite{repcut} simulation process.

\para{Differential exchange ~\cite{parendi}}  
In multithreaded or multicore simulations, only signals updated in the current cycle are communicated to other threads or cores for synchronization.
This selective propagation minimizes unnecessary data transfers and improves simulation efficiency.

\para{Instance reuse ~\cite{dedup}}  
To improve code reuse, prior work~\cite{dedup} identifies multiple instances of the same module within a digital circuit and generates shared code that can be applied to all such instances.
In \ourscheme, applying this technique generalizes the \emph{operator fusion} optimization (Section~\ref{subsec:impl-opt}) by constructing recurring, design-specific large operations.


\para{ML for RTL simulation ~\cite{MLmodel}} 
Incorporating a machine learning model to predict simulation results constitutes a cascade-level modification.
Prior work~\cite{fusemax} has shown that advanced ML architectures, such as attention mechanisms, can be expressed as a cascade of Einsums.

\subsection{Mapping-Level Optimizations}

\para{Hypergraph partitioning ~\cite{repcut, gem}}  
Partitioning is a classic loop transformation strategy. It is classified as a mapping-level optimization because it increases the number of ranks in the iteration space and, therefore, the available traversal orders and parallelizations.

\para{Traversal exploiting temporal locality ~\cite{dedup}} identifies duplicated instances and employs shared code to improve code reuse.
It further introduces a scheduling strategy that places tasks executing the shared code close together in time, thereby exploiting temporal locality. 

\para{Parallelization across partitions ~\cite{repcut, parendi, gem, Metro-Mpi}}
Once the dataflow graph is partitioned, each partition can be distributed across different threads to improve parallelism.  
TeAAL distinguishes between partitioning and parallelism (spatial or temporal) as separate optimizations; please refer to the TeAAL paper~\cite{teaal} for more details.

\subsection{Format-Level Optimizations}
\para{Optimizing memory layout ~\cite{repcut}}  
Optimizing the data layout and its storage format in memory is a classic format-level optimization.

\subsection{Binding-Level Optimizations}

\para{Branch hints ~\cite{essent}}  
Compiler-based branch hints do not change the logical execution order, but they do affect the physical instruction layout (e.g., place the likely path fall-through and place the unlikely path behind a jump).


\para{Software pipelining ~\cite{ToraRTL}}  
Software pipelining refers to the interleaving of tasks to minimize idle time of hardware units.  
It determines the timing and assignment of tasks to specific execution units.

\para{Multi-platform task scheduling ~\cite{cuda}}  
Prior work, such as~\cite{cuda}, studies scheduling tasks across multiple platforms, e.g., CPU and GPU, to improve performance.  
Similar to software pipelining, this optimization controls task assignment and timing to hardware units.

\subsection{Data-Level Optimizations}

\para{Replication for parallelism ~\cite{repcut, parendi, gem}}  
In RepCut-style hypergraph partitioning, certain portions of the dataflow graph are replicated to fully decouple partitions by eliminating intra-cycle dependencies.
These duplicated components modify the structure of the dataflow graph and are therefore classified as data-level optimizations.


\section{Cascade for RepCut Simulation}\label{app:repcut-einsum}
\begin{figure}[h]
\centering
\scriptsize
\begin{einsumbox}[box:repcut-cascade]{}

\[
\begin{aligned}
OI_{c, i, n, o, r_1, r_0, s_0} &= \layerinput_{c, i, r_1, r_0} \cdot \operationinputmask_{i, n, o, r_1, r_0, s_0} :: \textstyle\bigwedge \leftarrow (\rightarrow) \\
\layeroutput_{c, i, n, r_1, s_0} &= OI_{c, i, n, o, r_1, r_0, s_0} :: \textstyle\bigwedge op\_u[n] (\leftarrow) \textstyle\bigvee\ op\_r[n](\rightarrow) \\
\layeroutput\_sel_{c, i, n, o*, r_1, r_0, s_0} & = OI_{c, i, n, o, r_1, r_0, s_0} :: \textstyle\bigwedge \mathbbm{1} (\leftarrow) \lll \mathbbm{1}(op\_s[n]) \\
\end{aligned}
\]

\[
\scalebox{1}{$
\layerinput_{c, i+1, r_1, s_0} =
\left\{
\begin{array}{l}
\layeroutput_{c, i, n, r_1, s_0} :: \bigwedge \mathbbm{1} (\leftarrow) \bigvee\, ANY(\rightarrow), n \notin n\_sel \\
\layeroutput\_sel_{c, i, n, o, r_1, r_0, s_0} :: \bigwedge \mathbbm{1} (\leftarrow) \bigvee\, ANY(\rightarrow), n \in n\_sel
\end{array}
\right.
$}
\]
\[
\scalebox{1}{$
\diamond : i \equiv I
$}
\]
\[
\scalebox{1}{$
\layerinput_{c+1, o, s_1, s_0} = \layerinput_{c, I, r_1, r_0} \cdot RUM_{r_1, r_0, s_1, s_0} :: 
\bigwedge \leftarrow (\rightarrow)
$}
\]
\[
\scalebox{1}{$
\diamond : c \equiv C
$}
\]
\end{einsumbox}
\smallskip
\noindent\small
\textbf{Cascade 2: RepCut Simulation Einsum Cascade.}
\end{figure}

We present the complete cascade of Einsums for the RepCut~\cite{repcut} simulation kernel (Cascade~\ref{box:repcut-cascade}). 
This cascade is similar to the RTeAAL Sim cascade (Cascade~\ref{box:einsum}), with the primary difference being the final Einsum, which captures the register synchronization step specific to RepCut.

In RepCut, the dataflow graphs are partitioned into separate sectors, incurring some replication overhead. 
Certain registers are duplicated across multiple partitions, along with the operations that consume these registers as inputs. 
Each register is updated in only one partition; therefore, at the end of each cycle, the updated value must be propagated to all other partitions that also read this register. We refer to this as the \emph{synchronization step}.

The $RUM$ (Register Update Map) tensor encodes the mapping of register updates to reads.
For each register, $RUM$ specifies the partition where it is updated and the partitions where it is read.
At the end of each cycle, this map is used to propagate updated register values across the $LI$ tensors of the reading partitions.

\end{document}